\documentclass[10pt,twocolumn,letterpaper]{article}

\usepackage[pagenumbers]{wacv} % To force page numbers, e.g. for an arXiv version

\usepackage{graphicx}
\usepackage{amsmath}
\usepackage{amssymb}
\usepackage{booktabs}
\usepackage{graphics}
\usepackage{comment}

\usepackage[pagebackref,breaklinks,colorlinks]{hyperref}

\usepackage[capitalize]{cleveref}
\crefname{section}{Sec.}{Secs.}
\Crefname{section}{Section}{Sections}
\Crefname{table}{Table}{Tables}
\crefname{table}{Tab.}{Tabs.}

\def\wacvPaperID{1489} % *** Enter the WACV Paper ID here
\def\confName{WACV}
\def\confYear{2024}

\begin{document}

\title{Denoising diffusion models for high-resolution microscopy image restoration}

\author{Pamela Osuna-Vargas\\
Frankfurt Institute for Advanced Studies (FIAS)\\
and Department of Computer Science, \\Goethe University Frankfurt \\
{\tt\small osuna@fias.uni-frankfurt.de}
\and
Maren H. Wehrheim\\
Frankfurt Institute for Advanced Studies (FIAS)\\
and Department of Computer Science, \\Goethe University Frankfurt \\
{\tt\small wehrheim@fias.uni-frankfurt.de}
\and
Lucas Zinz\\
Department of Computer Science, \\Goethe University Frankfurt \\
{\tt\small s7005665@stud.uni-frankfurt.de}
\and
Johanna Rahm \\
Institute of Physical and Theoretical Chemistry, \\Goethe University Frankfurt \\
{\tt\small rahm@chemie.uni-frankfurt.de}
\and
Ashwin Balakrishnan  \\
Institute of Physical and Theoretical Chemistry, \\Goethe University Frankfurt \\
{\tt\small balakrishnan@chemie.uni-frankfurt.de}
\and
Alexandra Kaminer\\
Institute of Physical and Theoretical Chemistry, \\Goethe University Frankfurt \\
{\tt\small kaminer@chemie.uni-frankfurt.de}
\and
Mike Heilemann\\
Institute of Physical and Theoretical Chemistry, \\Goethe University Frankfurt \\
{\tt\small heilemann@chemie.uni-frankfurt.de}
\and
Matthias Kaschube \\
Frankfurt Institute for Advanced Studies (FIAS)\\
and Department of Computer Science, \\Goethe University Frankfurt \\
{\tt\small kaschube@fias.uni-frankfurt.de}
}
\maketitle

\begin{abstract}
   Advances in microscopy imaging enable researchers to visualize structures at the nanoscale level thereby unraveling intricate details of biological organization. However, challenges such as image noise, photobleaching of fluorophores, and low tolerability of biological samples to high light doses remain, restricting temporal resolutions and experiment durations. Reduced laser doses enable longer measurements at the cost of lower resolution and increased noise, which hinders accurate downstream analyses. Here we train a denoising diffusion probabilistic model (DDPM) to predict high-resolution images by conditioning the model on low-resolution information. Additionally, the probabilistic aspect of the DDPM allows for repeated generation of images that tend to further increase the signal-to-noise ratio. We show that our model achieves a performance that is better or similar to the previously best-performing methods, across four highly diverse datasets. Importantly, while any of the previous methods show competitive performance for some, but not all datasets, our method consistently achieves high performance across all four data sets, suggesting high generalizability. 
   % Perform at least as well as previous methods, but most importantly, it does so for all three datasets
\end{abstract}    
\section{Introduction}
\label{sec:intro}

% first mention the importance of microscopy.
High-resolution microscopy imaging impacts biology by revealing the detailed structure of living systems, providing a crucial basis for visualization, analysis, and interpretation. However, obtaining detailed microscopy images often requires immoderate imaging conditions, such as high light intensity. This can lead to photobleaching and phototoxicity, compromising the integrity of biological samples and limiting the duration over which observations can be made \cite{waldchen_light-induced_2015}. Reducing the light intensity during imaging can mitigate these effects, but often at the cost of higher noise, obscuring important details in the imaging data.

Denoising techniques have thus become essential in biological and medical microscopy applications, where preserving sample integrity while obtaining high-resolution images is often paramount. Important examples include resolving the temporal evolution of sub-cellular organization such as organelles, revealing their morphology, or identifying condition-dependent changes of these structures (e.g., stress or drugs). Traditional denoising methods, such as Gaussian filtering or wavelet transforms, often fall short in preserving fine details while removing noise. Recently, deep learning-based approaches have shown significant improvements in image-denoising tasks, also in the field of microscopy \cite{sun_pix2pix_2022, weigert_content-aware_2018, von_chamier_democratising_2021, zhang_image_2018, stringer_cellpose3_2024, gomez-de-mariscal_harnessing_2024, laine_imaging_2021, kefer_performance_2021}.

%Among these, Denoising Diffusion Probabilistic Models (DDPMs) have emerged as a powerful tool for generating high-fidelity images from noisy data.

% TODO: link the ideas
% whose principle relies on slowly adding random noise to data samples, and then learning the reverse process to reconstruct the samples from noise. T
Denoising Diffusion Probabilistic Models (DDPMs) \cite{ho_denoising_2020} are powerful models that generate images from noisy inputs by iteratively removing noise in a diffusion process. Motivated by their ability to generate fine-scale images, these models may be well-suited for denoising in biology, where reconstruction of detailed structures is crucial. Indeed, DDPMs have been recently shown to effectively remove signal-dependent and -independent noise in medical and biological imaging data \cite{lu_diffusion-based_2024, pan_diffuseirdiffusion_2023, liu_diffusion_2023, gao_corediff_2024, xia_low-dose_2022, gong_pet_2022, xiang_ddm2_2023}. However, it is currently unclear how well DDPMs perform in denoising fluorescence microscopy data, especially since the noise characteristics of such data can differ significantly from that of other data types \cite{belthangady_applications_2019}.

% DDPMs can handle various noise levels and types, thus making them promising candidate models for microscopy data denoising.

% \cite{lu_diffusion-based_2024} uses DDPMs to restore electron microscopy 
% \cite{pan_diffuseIR} DDPMs to restore EM

% either task is different \cite{waibel_diffusion_2023}, (2D->3D, generation of synthetic data \cite{saguy_this_2024}, latent space exploration on cryo-em data \cite{kreis_latent_2022})

Here we demonstrate that DDPMs can be highly effective in denoising a diverse range of fluorescence microscopy datasets, resolving fine structural details of the samples being studied. We use DDPMs to model the complex noise characteristics of different types of low-light microscopy images and use these models for high-quality image restoration, enabling longer imaging periods without sacrificing the sample quality. Additionally, we show that computing an average across several denoised reconstructions, exploiting the DDPM's stochasticity, can further enhance the performance significantly. We systematically test the denoising performance using four different datasets acquired through stimulated emission depletion (STED, datasets 1 and 2), confocal and Airyscan super-resolution (dataset 3), and single example and averaged confocal (dataset 4) fluorescence microscopy. These datasets vary considerably regarding the acquisition process, the samples being imaged (e.g., mitochondria or zebrafish), the sample condition (live vs fixed), the noise levels and structure, and the procedure for obtaining the low- and high-noise examples. This diversity of imaging conditions allows us to test the robustness of the proposed denoising method across different fluorescence microscopy applications. Importantly, our approach shows a performance that is higher than, or at least similar to current benchmark models across \textit{all} tested datasets, demonstrating its broad applicability and robustness. In fact, no other benchmark model performs consistently as high across datasets as our proposed DDPM. Overall, we find that our DDPMs architecture provides a highly competitive method for denoising fluorescence microscopy data, and integrating such models into the microscopy imaging workflow could pave the way for more accurate and less invasive imaging practices in the future.

Our contributions are as follows: 
\begin{itemize}
    \item We introduce a DDPM architecture for fluorescence microscopy image denoising that achieves competitive performance across diverse datasets.
    \item We suggest a repeated sampling scheme that increases the signal-to-noise ratio, building on the stochastic denoising process of the DDPM.
    \item We publish two novel, challenging denoising datasets containing STED images of fixed-cell microtubules and live-cell mitochondria. 
\end{itemize}

\section{Related work}
\label{sec:related_work}

Diffusion models have emerged as a powerful tool for several computer vision tasks \cite{nichol_improved_2021, ho_cascaded_2022, nichol_glide_2022, baranchuk_label-efficient_2022, amit_segdiff_2022, saharia_image_2021, saharia_palette_2022} showing greater training stability and superior image quality compared to previous generative models \cite{dhariwal_diffusion_2021, ho_denoising_2020, rombach_high-resolution_2022, saharia_palette_2022}. In the medical and biological domain DDPMs have been applied to segmentation \cite{wolleb_diffusion_2021, zbinden_stochastic_2023}, anomaly detection \cite{wolleb_diffusion_2022, lyu_conversion_2022}, image-to-image translation \cite{ozbey_unsupervised_2023}, molecule generation \cite{pao-huang_scalable_2023}, or 2/3D generation \cite{waibel_diffusion_2023}. Several studies have proposed using DDPMs in microscopy \cite{guo_diffusion_2023}, to, e.g., predict 3D cellular structure from 2D images \cite{waibel_diffusion_2023}, reconstruct 3D biomolecule structure in Cryo-EM data \cite{kreis_latent_2022}, generate super-resolution images \cite{saguy_this_2024}, or design drug molecules \cite{igashov_equivariant_2022}.

In recent years, deep learning methods have replaced classical denoising methods due to their better performance \cite{lehtinen_noise2noise_2018, weigert_content-aware_2018, isola_image--image_2018,krull_noise2void_2019, lequyer_fast_2022, chaudhary_fast_2022, ebrahimi_deep_2023, lu_emdiffuse_2023, qiao_zero-shot_2024, chaudhary_baikal_2024}. One popular self-supervised denoising method is Noise2Void \cite{krull_noise2void_2019}, where pixel-wise independent noise is assumed such that nearby pixels within a single example provide useful information for denoising. Pix2pix \cite{isola_image--image_2018} was introduced as a general-purpose framework for image-to-image translation tasks, using conditional generative adversarial networks (cGANs). 
%However, the method can also be extended to other problems such as computerized tomography (CT) scan image generation \cite{toda_lung_2022}, sketch-to-photograph conversion \cite{raghavendra_transfer_2022}, or depth image generation \cite{shimada_pix2pix-based_2022}. 
Several works extended pix2pix to image denoising tasks \cite{sun_pix2pix_2022, khan_crowd_2023, raposo_ultrasound_2021}. Additionally, the widely-used content-aware image restoration (CARE) network \cite{weigert_content-aware_2018} incorporates a U-Net architecture \cite{ronneberger_u-net_2015} to denoise low-resolution fluorescence data. More recently, the UNet-RCAN \cite{ebrahimi_deep_2023} first restores contextual features using a U-Net, and then leverages the ability of Residual Channel Attention Networks (RCAN) \cite{zhang_image_2018} to reconstruct super-resolution images. Diffusion models have also been used for denoising ultrasounds \cite{guha_sddpm_2023, asgariandehkordi_deep_2023}, CT or PET images \cite{liu_diffusion_2023, gao_corediff_2024, xia_low-dose_2022, gong_pet_2022}, MRI data \cite{xiang_ddm2_2023}, retinal images \cite{hu_unsupervised_2022}, or EM data \cite{lu_emdiffuse_2023}. Chaudhary et al.\cite{chaudhary_baikal_2024} recently proposed a DDPM for denoising fluorescence microscopy data using unpaired samples. However, while not relying on paired samples is advantageous, the performance of methods using paired samples is often higher \cite{weigert_content-aware_2018, ebrahimi_deep_2023}. Whereas all of the above-mentioned methods significantly enhance image quality, they still show various limitations, such as blurriness, hallucinations, low signal-to-noise ratio, or excessive smoothing of the sample. 

% Here, we introduce a conditioned DDPM to denoise fluorescence microscopy data using paired samples of low- and high-resolution images. 
% DDPMs have been applied for denoising several data types including EM, MRI, CT, and PET \cite{lu_diffusion-based_2024, pan_diffuseirdiffusion_2023, liu_diffusion_2023, gao_corediff_2024, xia_low-dose_2022, gong_pet_2022, xiang_ddm2_2023}. 
% https://github.com/CroitoruAlin/Diffusion-Models-in-Vision-A-Survey could use this for REFS

% Also, paired and unpaired methods often lead to complementary denoising behavior \cite

% % chaudhary_baikal_2024
%  Although Lu et al. \cite{lu_diffusion-based_2024, pan_diffuseir_20} elucidated the potential of diffusion models to restore microscopy data, their work was limited to electron microscopy (EM).
%  Although Lu et al. \cite{lu_diffusion-based_2024, pan_diffuseir_20} elucidated the potential of diffusion models to restore microscopy data, their work was limited to electron microscopy (EM).
\section{Methods}
\label{sec:methods}
\subsection{Data}
We here use several microscopy datasets to test the performance of DDPMs for image denoising. Specifically, we use two novel datasets containing stimulated emission depletion (STED) microscopy images of microtubules (immunostained for $\alpha$-tubulin) and of mitochondria that we publish alongside this paper. Further, we use two open-source datasets containing high- and low-resolution images of ex-vivo synapses \cite{xu_cross-modality_2023}, and confocal images of zebrafish embryos \cite{zhang_poisson-gaussian_2019}. These datasets differ across microscopy types, noise levels, sample types, and ground truth data generation, providing a challenging generalization task for denoising methods in fluorescence microscopy (see Suppl. Tables \ref{tab:dataset_description} and \ref{tab:dataset_diversity} for a comparison between datasets). 

\subsubsection{Fixed-cell microtubules and live-cell mitochondria datasets}
\label{sec:internal_dataset}
STED imaging of both fixed-cell microtubules (immunostained for $\alpha$-Tubulin) and live-cell mitochondria (stained with transient HaloTag ligand Hy4–SiR \cite{kompa_exchangeable_2023} for TOM20) were performed with an Abberior expert line microscope (Abberior Instruments, Germany). The setup uses an Olympus IX83 body (Olympus Deutschland GmbH, Germany) where the imaging was done using a UPLXAPO 60x NA 1.42 oil immersion objective (Olympus Deutschland GmbH, Germany). A 640 nm excitation laser was used to acquire sample images (both confocal and STED imaging). In the case of STED images depletion was performed with a 775 nm laser with a donut PSF (for planar and long-term imaging) with a delay of 750 ps and fluorescence photons between 750 ps and 8.75 ns were detected between every laser pulse. Both 640 nm and 775 nm lasers were pulsed at 40 MHz. Fluorescence was collected in the spectral range of 650 nm to 760 nm using an avalanche photodiode (APD). The pixel size used for the microtubule dataset was 25 nm and for the mitochondria dataset was 20 nm. The power levels of the lasers used and the parameters used for imaging are summarized in Table \ref{tab:data}. Note that the power levels for the lasers were measured at the back focal plane. Several samples were imaged in the case of both microtubules and mitochondria. The low-intensity images were measured by changing different parameters that influence the noise in a STED image, such as the excitation laser intensity, the number of lines integrated into the image, and the pixel dwell time. The depletion laser intensity was kept constant for the low- and high-intensity images to keep the resolution information intact.
% low-intensity STED timeseries of live-cell mitochondria with a time resolution of 4.6 s and imaged for 46 min.
One particularity of the microtubules and the mitochondria datasets is that, due to the difference in light dosage, the pixel distributions of low- and high-resolution images cover very distinct ranges. The low-resolution images contain only few pixel values, which poses an extra challenge for any denoising algorithm. 

\begin{table}[tb]
  \centering
  \resizebox{\columnwidth}{!}{%
  \begin{tabular}{c@{\hskip .05in}c@{\hskip .05in}c@{\hskip .05in}c@{\hskip .05in}c} 
    \toprule
     Intensity & \shortstack{Excitation \\(\textmu W)}  & \shortstack{Depletion \\(mW)} & \shortstack{Dwell time \\(\textmu s)}  & \shortstack{Light dosage to \\ low-intensity}  \\
    \midrule
    Low-tubulin & 1.5 & 60 & 1.5 & 1\\
    High-tubulin & 10.6 & 60 & 25 & 17\\
    Low-mitoc. & 1.5 & 174 & 2 & 1\\
    High-mitoc. & 8 & 174 & 20 & 10\\
  \bottomrule
  \end{tabular}%
  }
  \caption{\textbf{Measurement conditions of STED microtubules and mitochondria image dataset.} Note: Dwell times given are the total dwell time and take into account the number of line integrations.}
  \label{tab:data}
\end{table}
% $\frac{\text{light dosage}}{\text{low-intensity}}$

\subsubsection{Synapse dataset}
Additionally, we use the data published by Xu et al. \cite{xu_cross-modality_2023} containing low-resolution confocal images and high-resolution Airyscan imaging ground-truth (GT) from tissue slices of different cortical regions of transgenic mice. First, the high-resolution volumes were acquired immediately after the corresponding low-resolution images to reduce registration errors. The authors additionally curated the quality of the low-resolution images to replicate the image quality of \textit{in vivo} two-photon data. 
%  lateral resolution of 0.063 μm per px and an axial resolution of 0.33 μm per px

\subsubsection{Zebrafish dataset}
Finally, we use one of the partitions from the open-source Fluorescence Microscopy Denoising dataset \cite{zhang_poisson-gaussian_2019}. The partition we employ consists of confocal images of fixed zebrafish embryos [EGFP labeled Tg(sox10:megfp) zebrafish at 2 days post fertilization]. All animal studies were approved by the university’s Institutional Animal Care and Use Committee. 
Characteristic of this dataset is the noise type, shown to be Poisson-dominated due to its imaging modality. Each of the fields of view (FOVs) was captured 50 times, each exhibiting a different noise realization. Authors provide images with different noise levels, generated by averaging $S$ noisy raw images. With an increasing number of images used for averaging, the peak signal-to-noise ratio (PSNR) of the averaged images increases, making the denoising task more simple. Here we employ the most difficult case, using $S=1$ as raw images and $S=50$ as ground-truth images.

\subsection{Conditional denoising diffusion probabilistic models}
\label{sec: methods-ddpm}
We here follow the work proposed by Saharia et al. \cite{saharia_palette_2022} to adapt denoising diffusion probabilistic models (DDPMs) \cite{ho_denoising_2020} to a conditional image generation model.
 
Consider a data set of input-output (i.e. high-low noise) image pairs $(\textbf{x}_i, \textbf{y}_i)_{i=1}^N$ drawn from an unknown conditional distribution $p(\textbf{y}|\textbf{x})$. We aim to approximate $p(\textbf{y}|\textbf{x})$ using a stochastic iterative refinement process conditioned on a source image $\textbf{x}$ to generate a target image $\textbf{y}$.  Specifically, the conditioned DDPM is trained to generate a target image $\textbf{y}_0$ in $T$ steps, starting from an image of isotropic Gaussian noise $\textbf{y}_T \sim \mathcal{N}(0, \textbf{I})$. Via $T$ successive iterations $t$, the model computes $\textbf{y}_0 \sim p(\textbf{y} | \textbf{x})$ using learned conditional transition distributions $p_\theta (\textbf{y}_{t-1} | \textbf{y}_t, \textbf{x})$, where $\theta$ are the model parameters.

In the \textit{forward diffusion process} Gaussian noise is gradually added to the signal via a fixed Markov chain $q(\textbf{y}_t | \textbf{y}_{t-1})$. Specifically, by reparameterizing the variance and merging the Gaussian noise, we can sample $\textbf{y}_t$ at any step $t$ as:
\begin{equation}
\textbf{y}_t = \sqrt{\bar{\alpha}_t} \textbf{y}_0 + \sqrt{1-\bar{\alpha}_t}\boldsymbol{\epsilon}, \end{equation}
where $\boldsymbol{\epsilon} \sim \mathcal{N}(0, \textbf{\textit{I}})$, and $\bar{\alpha}_t = \prod_{i=1}^t \alpha_i \in (0,1)$ determines the variance of the noise added in each iteration, according to the variance schedule $\{\alpha_i\}^T_{i=1}$. 

Then, the above process is reversed by the \textit{reverse diffusion process}.
The conditional DDPM employed here uses a reverse Markov chain conditioned on $\textbf{x}$ to iteratively recover the signal $\textbf{y}_0$ from noise $\textbf{y}_T$. A denoising model $\boldsymbol{\epsilon}_\theta$ that follows from $p_\theta$ (as later described in Eqs. \ref{eq: p_theta}, \ref{eq: sigma_theta}, \ref{eq: mu_theta}) is trained to predict the noise $\boldsymbol{\epsilon}$ using the conditioning source image $\textbf{x}$, a noisy target image $\textbf{y}_t$, and additional conditioning on the statistics for the noise variance $\bar{\alpha}_t$ \cite{song_generative_2020, chen_wavegrad_2020}.

The learned inference process is defined as a conditional transition distribution $p_\theta(\textbf{y}_{t-1}|\textbf{y}_t, \textbf{x})$. The target image $\textbf{y}_0$ is then approximated by $\hat{\textbf{y}}_0$:

\begin{equation}
\hat{\textbf{y}}_0 = \frac{1}{\sqrt{\bar{\alpha}_t}} \left(\textbf{y}_t - \sqrt{1 - \bar{\alpha}_t} \boldsymbol{\epsilon}_\theta(\textbf{x}, \textbf{y}_t, \bar{\alpha}_t)\right).
\end{equation}

We optimize the parameters $\theta$ of the noise predictor model $\boldsymbol{\epsilon}_\theta$ by defining the learning objective $L$ as:

\begin{equation} 
\mathbb{E}_{(\textbf{x},\textbf{y}_0)} \mathbb{E}_{\boldsymbol{\epsilon}, \bar{\alpha}_t}\left[\big\|\boldsymbol{\epsilon}_\theta (\textbf{x}, \sqrt{\bar{\alpha}_t} \textbf{y}_0 +  \sqrt{1-\bar{\alpha}_t}\boldsymbol{\epsilon}, \bar{\alpha}_t) - \boldsymbol{\epsilon} \big\|_2^2\right],
\end{equation}
where $\boldsymbol{\epsilon} \sim \mathcal{N}(0, \textbf{\textit{I}})$, $(\textbf{x},\textbf{y}_0)$ is sampled from the training set, $\bar{\alpha}_t = \prod_{i=1}^t \alpha_i$ given $\alpha_i$ defined by the variance schedule $\{\alpha_i\}^T_{i=1}$ and $t$  uniformly sampled from $[1,T]$.
 
%$ \boldsymbol{\epsilon} \sim \mathcal{N} (\textbf{0}, \textbf{\textit{I}})$,  
% is it necessary to include this?

Recall that our goal is to learn the conditioned transition distributions $p_\theta (\textbf{y}_{t-1}|\textbf{y}_t)$ in the reverse diffusion process:
\begin{equation}
\label{eq: p_theta}
p_\theta (\textbf{y}_{t-1}|\textbf{y}_t)= \mathcal{N} \left(\textbf{y}_{t-1}; \mu_\theta (\textbf{x}, \textbf{y}_t, \bar{\alpha}_t), \Sigma_\theta(\textbf{x}, \textbf{y}_{t}, \bar{\alpha}_t)\right).
\end{equation}
Instead of learning the diagonal variance $\Sigma_\theta$, we fix it as proposed by \cite{ho_denoising_2020}: 
\begin{equation}
\label{eq: sigma_theta}
\Sigma_\theta (\textbf{x}, \textbf{y}_{t}, \bar{\alpha}_t) = (1-\bar{\alpha}_t) \textbf{\textit{I}},
\end{equation}
and parametrize the mean $\mu_\theta$ as:

\begin{equation}
\label{eq: mu_theta}
\mu_\theta (\textbf{x}, \textbf{y}_t, \bar{\alpha}_t) = \frac{1}{\sqrt{\alpha_t}}\left(\textbf{y}_t - \frac{1 - \alpha_t}{\sqrt{1 - \bar{\alpha}_t}} \boldsymbol{\epsilon}_\theta(\textbf{x}, \textbf{y}_t, \bar{\alpha}_t)\right),
\end{equation}
which together allows for an iterative refinement in the following form:

\begin{equation}
\textbf{y}_{t-1} \leftarrow  \frac{1}{\sqrt{\alpha_t}}\left(\textbf{y}_t - \frac{1 - \alpha_t}{\sqrt{1 - \bar{\alpha}_t}} \boldsymbol{\epsilon}_\theta(\textbf{x}, \textbf{y}_t, \bar{\alpha}_t)\right) + \sqrt{1 - \alpha_t} \boldsymbol{\epsilon}.
\end{equation}

\subsection{Model architecture}
The underlying structure of the DDPM is a U-Net \cite{ronneberger_u-net_2015}. We adjust the conditional DDPM architecture of Saharia et al. \cite{saharia_palette_2022} to make the model more robust to different types of fluoresence microscopy data. Our changes are inspired by recently proposed improvements \cite{karras_analyzing_2024}, which identified weaknesses in the training dynamics of the traditional ADM architecture \cite{dhariwal_diffusion_2021}. Every resolution level of the U-Net consists of two blocks with convolutional layers for downsampling and transposed convolutional layers for upsampling, followed by self-attention at resolution 32 with 32 heads. The residual branch uses two convolutional layers, each preceded by a SiLU nonlinearity (see Sec. \ref{SI-architecture} and Fig. \ref{fig: SI-supp-architecture}). Foremost,  all operations, such as convolutions, activations, concatenation, and summation, are modified such that the expectation value of their magnitudes is preserved.
% network initialization using Kaiming method (He et al. 2015, ICCV)

Each denoising step is conditioned by the noise level information, which is encoded by an auxiliary embedding network into Fourier features by applying random frequencies and phases to the noise level information  (see Sec. \ref{SI-architecture} and Fig. \ref{fig: SI-supp-architecture} for more details), as opposed to the ADM's positional embedding scheme that employs a sinusoidal encoding \cite{dhariwal_diffusion_2021}. To condition on low-resolution images, each denoising step receives as input a concatenation of the conditioning low-resolution image $\textbf{x}$ and the prediction $\textbf{y}_t$ of the current time step $t$ (see Sec. \ref{sec: methods-ddpm}), sampled from a zero-mean isotropic Gaussian distribution at the first time step.

\subsection{Model training and evaluation}
We used image flipping, rotation, and Gaussian filtering as data augmentation techniques for training the DDPM. We trained the DDPM with AdamW \cite{loshchilov_decoupled_2019} optimizer, a batch size of 8, an initial learning rate of $2\mathrm{e-}4$, and use a cosine annealing schedule to adjust the learning rate during training. The epoch with the lowest mean absolute error (MAE) in the validation set is used to select the best epoch for testing. We used a cosine-based variance schedule \cite{nichol_improved_2021} with an offset $s$ = $8\mathrm{e-}3$, and set $T = 200$. All DDPMs were implemented in PyTorch \cite{paszke_automatic_2017} and trained with one GPU NVIDIA A100.

During inference, we utilized the stochasticity of the DDPMs and generated several predictions using the same condition input but different initial noise inputs. We then averaged across the predictions to remove any random noise not removed by the DDPM. We henceforth refer to the averaging of DDPM predictions as DDPM-avg.

\subsection{Quality control metrics}
We compute the mean absolute error (MAE), the peak signal-to-noise ratio (PSNR), the multiscale structural similarity index measure (MS-SSIM), and the learned perceptual image patch similarity (LPIPS) between each ground truth image and reconstruction (see Supplement for detailed description). Whereas the MAE measures the difference in pixel intensities, the PSNR quantifies the logarithmic peak error. The MS-SSIM additionally assesses the luminance, contrast, and structural information. LPIPS approximates the perceptual similarity between two images as would be indicated by humans. Additionally, we report the Pearson correlation coefficient, the resolution, and the normalized root mean square error (NRMSE) in the Supplement.

\subsection{Benchmarks}
We benchmark the performance of the conditioned DDPM against commonly used methods: Noise2Void (N2V) \cite{krull_noise2void_2019}, pix2pix \cite{sun_pix2pix_2022}, UNet-RCAN \cite{ebrahimi_deep_2023}, and CARE \cite{weigert_content-aware_2018}. For CARE and N2V probabilistic versions exist. However, we did not find a performance increase, even when averaging multiple output instances and thus report only results using the non-probabilistic versions.

The microtubules dataset contains very different pixel distributions between the low- and high-intensity images. To adjust for this effect it was sufficient for pix2pix and the DDPMs to clip the pixel values of the reconstruction to the range $[0,255]$, and cast the result to 8-bit format. However, we observed a significant difference between the predicted and ground truth pixel distribution of high-resolution images for N2V, CARE, and UNet-RCAN, which strongly affected performance. Therefore, for these models, we first clipped the pixel values to the range $[0,255]$, and then rescaled them to the ground truth pixel distribution of the training set using a linear transformation.

\section{Results \& Experiments}
\label{sec:results}
We use conditioned denoising diffusion probabilistic models (DDPM) to denoise different types of microscopy datasets. Specifically, we test our method using i) a novel dataset of low- and high-intensity STED images of fixed microtubules, ii) a second novel dataset of STED images of living mitochondria, iii) a publicly available dataset of synapses in mouse brain acquired with low- and high-resolution microscopes \cite{xu_cross-modality_2023}, and iv) another publicly available dataset of zebrafish imaged with confocal microscopy and different noise levels \cite{zhang_poisson-gaussian_2019}. We report results for a single generated prediction using the DDPM as well as an average across 15 such generated predictions (DDPM-avg). We benchmark the performance of our model to several previous methods (Noise2Void \cite{krull_noise2void_2019}, pix2pix \cite{isola_image--image_2018}, UNet-RCAN \cite{ebrahimi_deep_2023}, and CARE \cite{weigert_content-aware_2018}) and compare the performance between methods using the MAE, PSNR, MS-SSIM, and LPIPS (Figs. \ref{fig:results-tubulin}A - \ref{fig:results-zebrafish}A, Tables \ref{tab:metrics-tubulin}- \ref{tab:metrics-zebrafish}) in the main text. In the Supplement, we additionally report the NRMSE, resolution, and Pearson correlation.

% We benefited from the stochasticity of DDPMs to average multiple outputs given the same condition input, which improves the PSNR and NRMSE \ref{fig:effect-avg}. % mention the # of epochs for each dataset later 

\subsection{Denoising STED images of microtubules}
\begin{figure}[t]
  \centering
\includegraphics[width=0.5\textwidth]{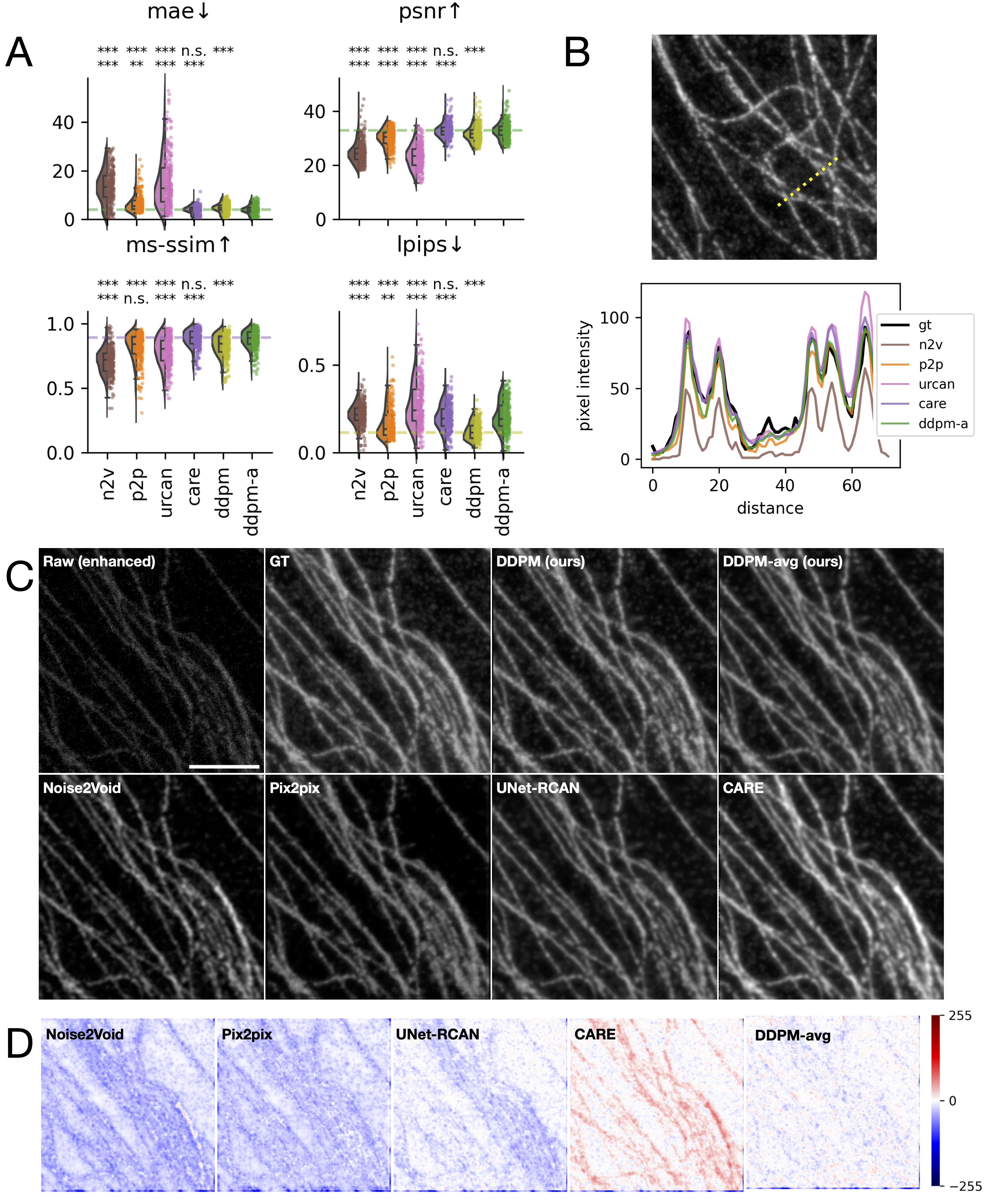}
  \caption{\textbf{Conditioned DDPMs outperform several previous methods in denoising STED images}. \textbf{A}) Performance comparison based on several evaluation metrics between our proposed method (DDPM and DDPM-avg) and several previously proposed benchmark models (see Methods for description of models and metrics). We indicate the median of the best-performing method for each metric as a dashed line in the respective color. Mood's median test was used to compute statistical significance; ***: $p < .001$, **: $p < .01$, *: $p < .05$, otherwise not significant (top (resp. bottom) row: difference to DDPM-avg (resp. DDPM); see Supplement for \textit{p}-values). Arrows indicate whether high or low values are optimal.  \textbf{B}) Pixel intensity profiles along the dashed yellow line (left) for all models (right).
  \textbf{C}) Top row: A representative low-intensity image (Raw) from the test dataset, the corresponding high-resolution version (ground truth, GT), and the results of our proposed method. Bottom row: results of the benchmark models. The scale bar indicates 2 \textmu m.
  \textbf{D}) Pixel-wise difference between the ground truth and the reconstruction for each model using the sample in C. Blue (red) values indicate a lower (higher) predicted pixel value.}
  \label{fig:results-tubulin}
\end{figure}

First, we train the conditioned DDPM to denoise low-intensity STED images of fixed microtubules. Despite aligning the low- and high-resolution image pairs (as described in Sec. \ref{sec:data_preprocessing}) all models predict a small shift in the reconstruction indicated by an offset in the peaks of the signal; thus we re-aligned the predictions to the GT for all models. Note that a highly accurate alignment is needed to compute a valid pixel-wise loss but can be omitted during practice. We observe that the DDPM accurately learns the pixel distribution of the target images and reconstructs the microtubule structures (Fig. \ref{fig:results-tubulin}A light and dark green elements). The DDPM and DDPM-avg outperform several previous methods (Noise2Void, pix2pix, and UNet-RCAN) in all evaluation metrics ($p < .001$ using Mood's median test). In addition, DDPM-avg achieves a similar performance to CARE for all evaluation metrics ($p > .43$, see Supplement). The signal profiles of the prediction align well with the ground truth signal profile across all models (Fig. \ref{fig:results-tubulin}B). In particular, the DDPM and pix2pix most closely preserve the peaks and troughs, whereas Noise2Void (brown) underpredicts the pixel intensities, and CARE (purple) overpredicts the peaks of the data that correspond to microtubule structures. This effect is also visible in the reconstructed images (Fig. \ref{fig:results-tubulin}C), where especially CARE exhibits very bright microtubules. The DDPM preserves the fine structures between the long microtubule structures that are removed by pix2pix and Noise2Void, which can be problematic when the imaged structures are small or lie orthogonal to the imaging plane. Further, the structures denoised with the DDPM are more sharp, when averaged across several examples compared to the single example (Fig. \ref{fig:results-tubulin}C top row last two columns), which is also reflected by the significant increase in performance when averaging across an increasing number of denoised examples (Fig. \ref{SI-effect-avg-all-metrics}). As illustrated by the difference maps in Fig. \ref{fig:results-tubulin}D, the errors in the predicted pixel intensities of the DDPM-avg are small and not systematic along the microtubule structures. In contrast,  all benchmark models show correlated errors along the microtubules, indicated by the pronounced red (overprediction) and blue (underprediction) lines. %
% Write about what we find? or in the Discussion?

\subsection{Denoising STED images of live-cell mitochondria}
\begin{figure}[t]
  \centering
\includegraphics[width=0.5\textwidth]{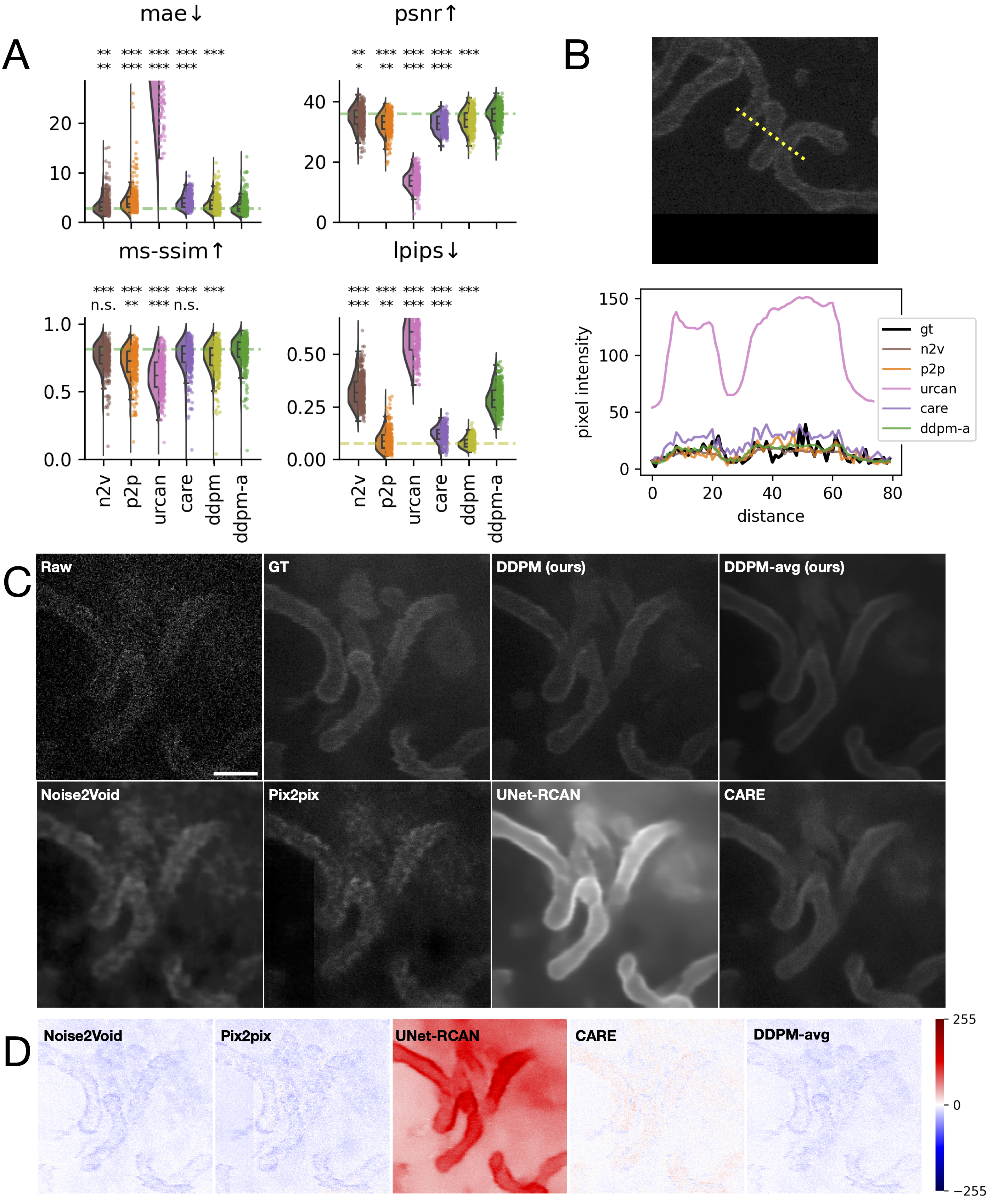}
  \caption{\textbf{Conditioned DDPMs outperform all benchmark models in denoising STED images of live-cell mitochondria}. \textbf{A} - \textbf{D} are as in Fig. \ref{fig:results-tubulin}, but for the live-cell mitochondria dataset. The scale bar in (\textbf{C}) indicates 1 \textmu m.}
  \label{fig:results-mitochondria}
\end{figure}

An important application of denoising in microscopy is live-cell imaging, as strong light exposure and photobleaching can strongly impact molecular biological processes. As one application, we therefore generate a dataset containing low- and high-resolution image pairs of living mitochondria (see Methods Sec. \ref{sec:internal_dataset}) and train the DDPM to predict the high-resolution data. Our proposed DDPM-avg achieves the highest performance across all metrics and only weakly underpredicts the true pixel distribution (Fig. \ref{fig:results-mitochondria}A-D, Table \ref{tab:metrics-mitochondria}). Note that only DDPM, CARE, and UNet-RCAN are able to reconstruct the mitochondria's outer membrane, but also predict structures to be smoother than in the GT. The predictions of Noise2Void and pix2pix are pixelated and often fail to enhance the mitochondria structures compared to the background. Note that all models fail to make predictions that are biologically fully plausible, as evident, for instance, by the 'open' membranes of some mitochondria. Again, CARE achieves the most similar results to the DDPM, both visually and based on the metrics, and the UNet-RCAN overpredicts the pixel intensities.
Interestingly, Noise2Void achieves very high performance across several evaluation metrics, but visually shows a poor performance.Averaging across several reconstructions again increases the performance in most metrics, except for LPIPS, where a single prediction of the DDPM achieves the highest performance (Figs. \ref{fig:results-mitochondria}A, \ref{SI-effect-avg-all-metrics}).

\begin{table}[tb]
  \centering
  \begin{center}
  \resizebox{\columnwidth}{!}{%
  \begin{tabular}{c@{\hskip .1in}c@{\hskip .1in}c@{\hskip .1in}c@{\hskip .1in}c@{\hskip .1in}c} 
    \toprule
     Model & MAE $\downarrow$ & PSNR (dB)$\uparrow$ & MS-SSIM $\uparrow$ & LPIPS $\downarrow$ \\
    \midrule
    Raw & $17.6$ & $20.88$ & $0.10$ & $0.71$ \\
    Noise2Void  & $13.26$  & $ 24.32 $  & $0.72$  & $0.22$ \\
    Pix2pix  & $5.38$  & $30.48$  & $0.84$  & $0.14$\\
    UNet-RCAN  & $12.73$  & $23.40 $  & $0.80$  & $0.24$\\
    CARE  & $4.11$  & $32.58 $  & $\textbf{0.89}$  & $0.20$\\
    \hline
    \textbf{DDPM}  & $4.71$ & $31.61$ & $0.84$ & $\textbf{0.11}$ \\
    \textbf{DDPM-avg} & $\textbf{3.99}$ & $\textbf{32.81}$ & $\textbf{0.89}$  & $0.19$ \\ 
  \bottomrule
  \end{tabular}%
  }
  \end{center}
  \caption{\textbf{Benchmarking the conditioned DDPM with the STED fixed-cell microtubule data set.} We report the median value of several performance metrics (MAE, PSNR, MS-SSIM, LPIPS) across models. See Supplement for the results with additional metrics (NRMSE, correlation).}
  \label{tab:metrics-tubulin}
\end{table}

\begin{table}[tb]
  \centering
  \begin{center}
  \resizebox{\columnwidth}{!}{%
  \begin{tabular}{c@{\hskip .1in}c@{\hskip .1in}c@{\hskip .1in}c@{\hskip .1in}c@{\hskip .1in}c} 
    \toprule
     Model & MAE $\downarrow$ & PSNR (dB)$\uparrow$ & MS-SSIM $\uparrow$ & LPIPS $\downarrow$ \\
    \midrule
    Raw & $10.12$ & $25.29$ & $0.25$ & $0.42$ \\
    Noise2Void  & $3.06$ & $34.78$  & $0.77$  &  $0.32$\\
    Pix2pix  & $3.76$  & $33.2$  & $0.73$  & $\textbf{0.08}$\\
    UNet-RCAN  & $39.25$  & $13.81$  & $0.62$  & $0.61$\\
    CARE  & $3.79$ & $32.79$  & $0.78$  & $ 0.12 $ \\
    \hline
    \textbf{DDPM}  &  $3.33$ & $34.1$ & $0.77$ & $\textbf{0.08}$ \\
    \textbf{DDPM-avg}  & $\textbf{2.72}$ & $\textbf{35.88}$ & $\textbf{0.81}$ & $0.28$ \\
  \bottomrule
  \end{tabular}%
  }
\end{center}
  \caption{\textbf{Benchmarking the conditioned DDPM on the live-cell mitochondria dataset.} See Supplement for the results with additional metrics.}
  \label{tab:metrics-mitochondria}
\end{table}

\subsection{Denoising microscopy images of synapses in the mouse brain}
\begin{figure}[t]
  \centering
\includegraphics[width=0.5\textwidth]{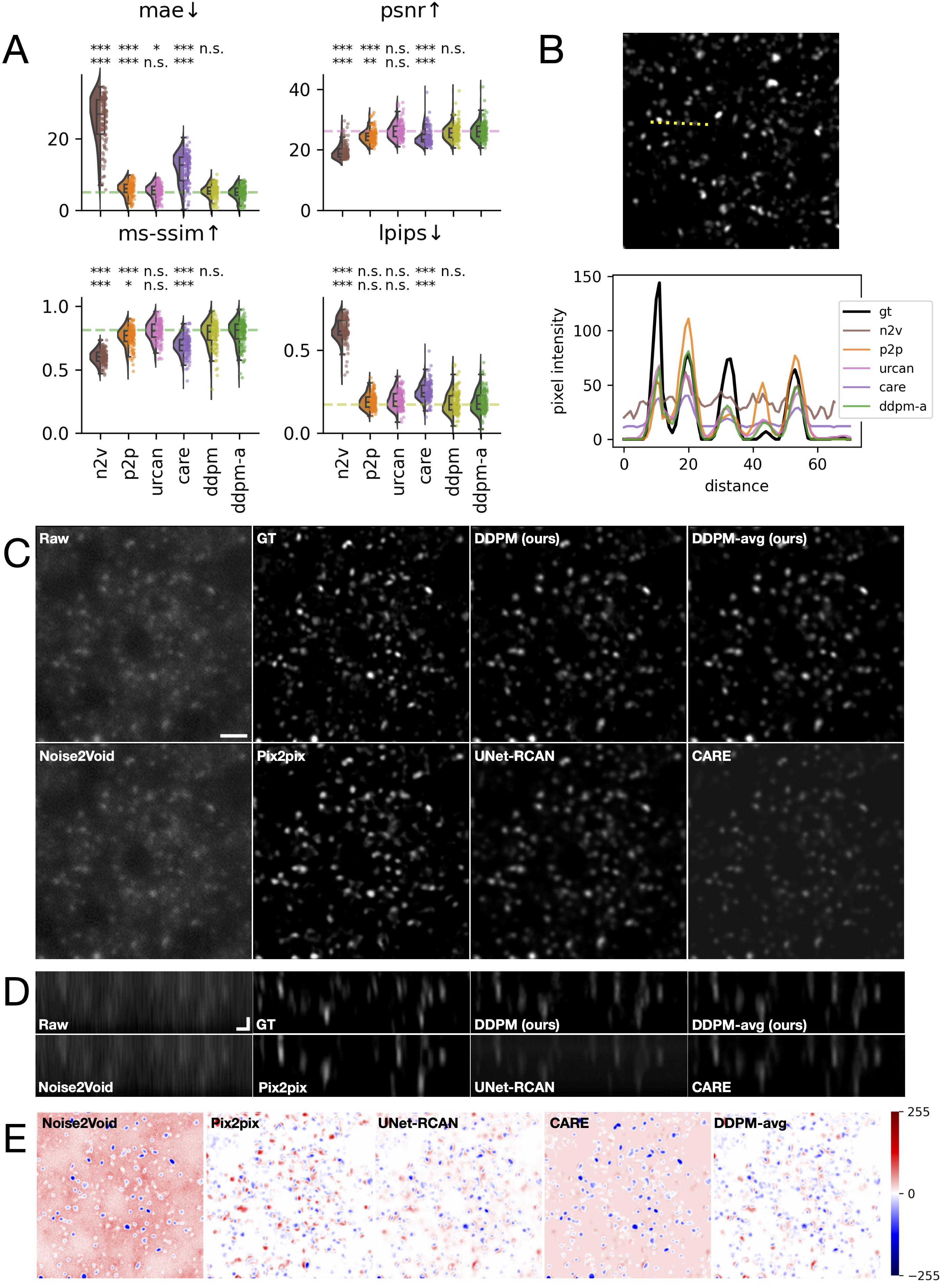}
  \caption{\textbf{Conditioned DDPMs outperform several previous methods in denoising confocal images of synapses in the mouse brain}. (\textbf{A}), (\textbf{B}), (\textbf{C}), (\textbf{E}) are as in Fig. \ref{fig:results-tubulin} (\textbf{A}), (\textbf{B}), (\textbf{C}), (\textbf{D}), respectively, but for the synapse dataset. \textbf{D}) Shows the $xz$-view of a sample. The scale bar indicates 1 \textmu m.}
  \label{fig:results-synapse}
\end{figure}

Next, we tested our model on confocal and super-resolution images of synapses in the mouse brain. Here, the model doesn't only have to denoise the image but also has to predict a signal across microscopy types, i.e., from confocal to super-resolution Airyscan quality. Our proposed conditioned DDPM successfully reconstructs the synaptic structures similarly to or better than previous methods (Fig. \ref{fig:results-synapse}A, Table \ref{tab:metrics-synapse}). Whereas for the tubulin dataset CARE shows the most similar results to the DDPM, here only UNet-RCAN achieves similar performance. Most methods predict blurry synapses with weak signal strengths for the bright synapses (see Fig. \ref{fig:results-synapse}B-E), indicating a high level of prediction uncertainty (see also uncertainty maps of the DDPM in Fig. \ref{fig: SI-uncertainy-maps}). From the predictions and error maps (Fig. \ref{fig:results-synapse}C-E) we observe a clear improvement of the DDPM in the background prediction with respect to CARE and Noise2Void; plus lower pixel error in contrast to pix2pix, and similar errors to UNet-RCAN in the restored intensities of the synapses (Fig. \ref{fig:results-synapse}E). Interestingly, averaging across several reconstructions does not improve the performance significantly in this data set (Figs. \ref{fig:results-synapse}A, \ref{SI-effect-avg-all-metrics}). Note that, in contrast to the error maps from the other datasets, all models struggle to accurately predict the synapse and/or background pixel values. This suggests further refinement is required in applications studying the changes in synapse strength and size, which is a task of central interest in neuroscience. 
% Note however, that the denoised synapses are sharper after averaging and that the large, well predicted background influences the computed metrics.

% Care doesn't resolve background
% n2v predicts in line profile almost straight line
% Now a lot of models underpredict the peaks, whereas only pix2pix shows slight overprediction
% only data set where the performance of ddpm-avg does not increase in all metrics
% synapses not sharp enough predicted, which can be seen by the red outlines around neurons in difference maps, whereas the centers of the neuron are underpredicted

\begin{table}[tb]
  \centering
  \begin{center}
  \resizebox{\columnwidth}{!}{%
  \begin{tabular}{c@{\hskip .1in}c@{\hskip .1in}c@{\hskip .1in}c@{\hskip .1in}c@{\hskip .1in}c} 
    \toprule
     Model & MAE $\downarrow$ & PSNR (dB)$\uparrow$ & MS-SSIM $\uparrow$ & LPIPS $\downarrow$ \\
    \midrule
    Raw & $27.35$ & $18.61$ & $0.60$ & $0.62$ \\
    Noise2Void  & $26.87$ & $ 18.74 $  & $0.60$  &  $0.61$\\
    Pix2pix  & $6.16$  & $ 24.43 $  & $0.77$  & $0.19$\\
    UNet-RCAN  & $5.66$  & $\textbf{26.11}$  & $\textbf{0.81}$  & $0.19$\\
    CARE  & $12.67$ & $23.35$  & $0.69$  & $ 0.24 $ \\
    \hline
    \textbf{DDPM}  &  $5.45$ & $25.61$ & $0.80$ & $\textbf{0.17}$ \\
    \textbf{DDPM-avg}  & $\textbf{5.09}$ & $25.96$ & $\textbf{0.81}$ & $0.18$ \\
  \bottomrule
  \end{tabular}%
  }
\end{center}
  \caption{\textbf{Benchmarking the conditioned DDPM on the synapse dataset.} See Fig. \ref{fig:SI-supp-metrics} and Tables \ref{tab:supp-metrics-internal}, \ref{tab:supp-metrics-external} for the results with additional metrics.}
  \label{tab:metrics-synapse}
\end{table}

% we make sure that images from one same volume are not in different datasets 
\subsection{Denoising confocal images of zebrafish embryos}
\begin{figure}[t]
  \centering
\includegraphics[width=0.5\textwidth]{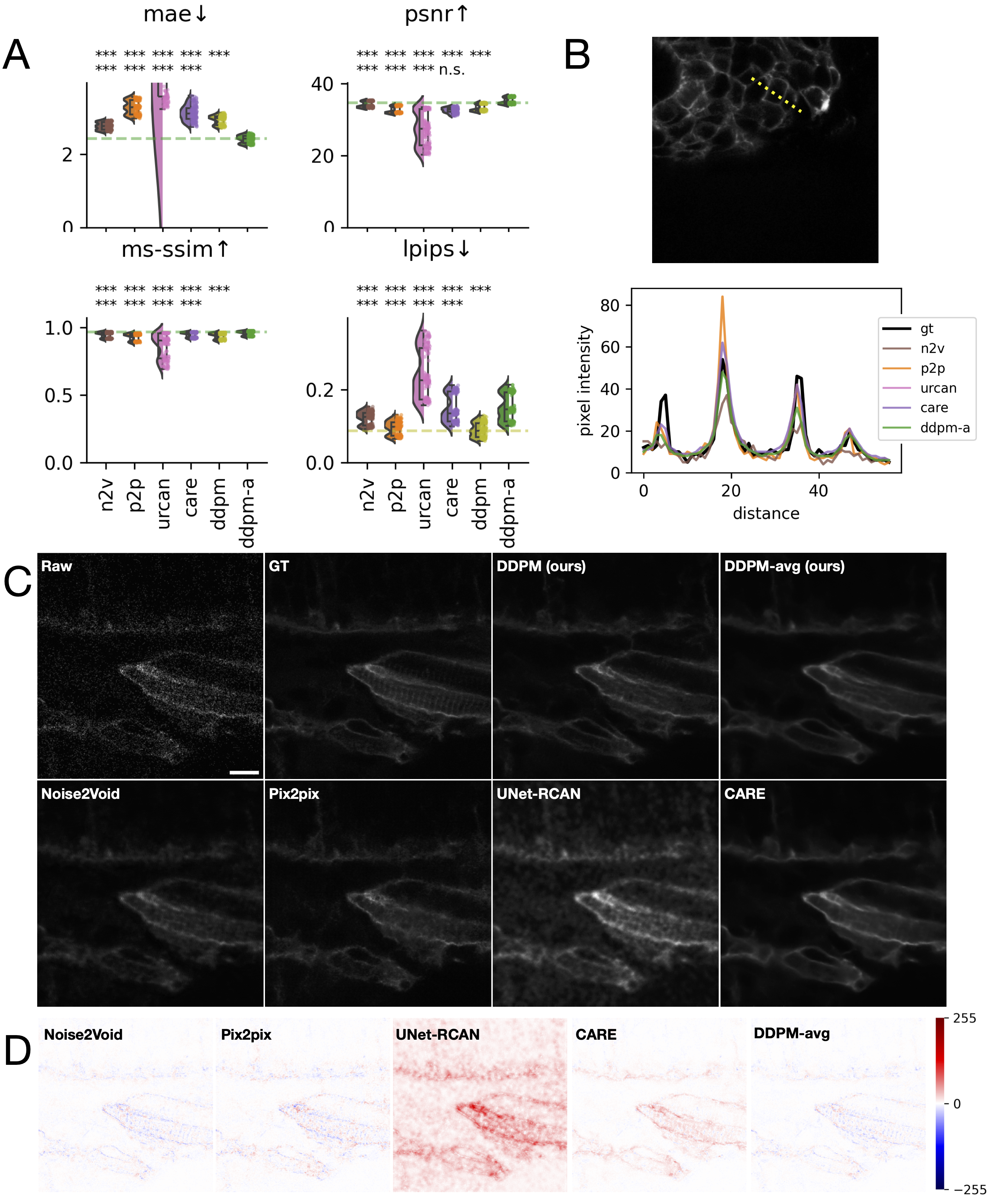}
  \caption{\textbf{Conditioned DDPMs outperform previous methods in denoising confocal images of zebrafish embryos.} (\textbf{A}), (\textbf{B}), (\textbf{C}), (\textbf{D}) are as in Fig. \ref{fig:results-tubulin}, but for the zebrafish dataset. The scale bar in (\textbf{C}) indicates 10 \textmu m.}
  \label{fig:results-zebrafish}
\end{figure}

Additionally, we trained our model to denoise confocal images of zebrafish embryos. The DDPM-avg outperforms all benchmark models in MAE, PSNR, and MS-SSIM, whereas the DDPM is best for LPIPS (Fig. \ref{fig:results-zebrafish}A, Table \ref{tab:metrics-zebrafish}). In this dataset the demarcation of sample structure and background is good for all methods, except the UNet-RCAN (Fig. \ref{fig:results-zebrafish}B-D). The averaging across several reconstructions eliminates noisy elements but also fine-grained structures in the data which are challenging for all models. Nevertheless, for this particular dataset, averaging implied a major improvement in the performance, except for the LPIPS, which might be a reflection of the higher smoothing (Figs. \ref{fig:results-zebrafish}A, \ref{SI-effect-avg-all-metrics}). The state-of-the-art denoising performance indicates that DDPMs effectively generalize to Poisson noise in microscopy data.

\begin{table}[tb]
  \centering
  \begin{center}
  \resizebox{\columnwidth}{!}{%
  \begin{tabular}{c@{\hskip .1in}c@{\hskip .1in}c@{\hskip .1in}c@{\hskip .1in}c@{\hskip .1in}c} 
    \toprule
     Model & MAE $\downarrow$ & PSNR (dB)$\uparrow$ & MS-SSIM $\uparrow$ & LPIPS $\downarrow$ \\
    \midrule
    Raw & $8.43$ & $25.28$ & $0.86$ & $0.52$ \\
    % \hline
    Noise2Void  & $2.76$ & $ 33.72 $  & $ 0.96 $  & $0.13$ \\
    Pix2pix  & $3.27$  & $ 32.01 $  & $0.95 $  & $0.10$ \\
    UNet-RCAN  & $7.44$ & $ 27.45 $ & $ 0.90 $  & $0.23$ \\
    CARE  & $3.10$ & $ 32.74 $ & $ 0.96 $ & $0.14$ \\
    \hline
    \textbf{DDPM}  & $2.98$ & $ 32.62 $ & $ 0.95 $ & $\textbf{0.09}$ \\
    \textbf{DDPM-avg}  & $\textbf{2.43}$ & $ \textbf{34.62} $ & $ \textbf{0.97} $ & $0.15$ \\
  \bottomrule
  \end{tabular}%
}
\end{center}
\caption{\textbf{Benchmarking the conditioned DDPM with the confocal zebrafish dataset.} See Supplement for the results with additional metrics.}
\label{tab:metrics-zebrafish}
\end{table}

\begin{comment}
\subsection{Robustness}
\begin{itemize}
    \item Hallucination?
    \item 
\end{itemize}
\end{comment}
\section{Discussion}
\label{sec:discussion}
In order to reduce phototoxicity while maintaining image quality in fluorescence microscopy, we explore the potential of image-conditioned diffusion models to denoise microscopy data in different datasets. We show the effectiveness of DDPMs in restoring (i) microtubule structures, (ii) living mitochondria (both acquired with STED microscopy), (iii) synaptic structures (confocal and high-resolution Airyscan imaging), and (iv) zebrafish embryos (confocal imaging). 

The stochastic property of DDPMs allows to repeatedly generate slightly different predictions from a single low-resolution input. We leverage this property by estimating averaged reconstructions from 15 individual predictions which increases the performance for three datasets significantly (microtubules, mitochondria, and zebrafish). Further, computing uncertainty maps from several predictions highlights the pixel-wise prediction uncertainty of the DDPM.

We compare the performance of the DDPM with several benchmark models: Noise2Void, pix2pix, UNet-RCAN, and CARE; all of which are based on supervised learning, except for Noise2Void. We find that for all four datasets, our proposed DDPMs achieve similarly high and sometimes higher performance than the best-performing benchmark model. CARE, UNet-RCAN, and the DDPM denoise the structural features well enough for further biological inspection. The DDPM-avg best predicts the pixel intensities across datasets, followed by CARE or UNet-RCAN. However, compared to the DDPM the predictions of CARE are often more blurry and the UNet-RCAN tends to overestimate the pixel intensities. In general, the DDPM not only denoises the sample structures more precisely but also better demarcates the background from the sample. Importantly, the DDPM is consistently among the best (top two) performing models across all tested datasets, whereas the performance of the benchmark models varies considerably across datasets. These results suggest a high degree of generalizability of DDPMs for denoising fluorescence microscopy data across different biological systems and microscopy conditions. 

We compare the performance of all models using several different performance metrics. The DDPM-avg is always the best-performing model for the MAE and MS-SSIM and at least second best for PSNR. LPIPS is always higher for single than for averaged reconstructions with the DDPM, suggesting that averaging reduces the perceptual similarity. 

%One limitation of our study is that, 
We here test several microscopy techniques and four widely used denoising methods as benchmarks. We acknowledge that other important microscopy techniques (e.g. widefield imaging, structured illumination microscopy) and denoising methods (e.g., ZS-DeconvNet \cite{qiao_zero-shot_2024}, Noise2Fast \cite{lequyer_fast_2022 }) exist. 

While DDPMs achieve the highest performance across several datasets, the costs of training and application are higher than for the other benchmark models. 
Future work could test the performance of the different denoising methods on biologically relevant downstream tasks, such as synapse detection or tracking and quantification of changes in fluorescence intensity, morphology, or motility of cell organelles across time.

% , or quantifying mitochondrion motility and morphological features.

\section{Conclusion}
In this work, we explore the applicability of DDPMs to denoise fluorescence microscopy images across different imaging modalities, ranging from STED and Airyscan to confocal microscopy, and a diversity of conditions, such as light-dosage, and cross-modality acquisition. We leverage recent improvements in the architecture of DDPMs to obtain a highly reliable training process. Also, we further increase the performance of the DDPM by averaging multiple high-resolution predictions conditioned on the same low-resolution image, thereby leveraging the stochastic nature of DDPMs. Overall, DDPMs perform better or similarly to current state-of-the-art denoising approaches. Foremost, their effectiveness is consistently high across the four datasets we test, suggesting this method is broadly applicable to a wide range of microscopy imaging data.
\label{sec:conclusion}

\clearpage
{

%\pagebreak
%%%%%%%%% REFERENCES
{\small
\bibliographystyle{ieee_fullname}
\bibliography{references}
}

\clearpage
\setcounter{page}{1}
\setcounter{section}{0}
\setcounter{figure}{0}
\setcounter{table}{0}
% \maketitlesupplementary

%\renewcommand \thesection{S\section}
%\renewcommand\thetable{S\table}
%\renewcommand \thefigure{S\figure}
\renewcommand{\thefigure}{S\arabic{figure}}
\renewcommand{\thesection}{S\arabic{section}}
\renewcommand{\thetable}{S\arabic{table}}

\twocolumn[
\begin{center}
    \Large \textbf{Supplemental Materials: Denoising diffusion models for high-resolution microscopy image restoration}
\end{center}
\vspace{1em} 
]

\section{Datasets}

\subsection{Data pre-processing}
\label{sec:data_preprocessing}
For the microtubules and synapse datasets, we used the itk library v5.4rc1 \cite{mccormick_itk_2014} to rigidly align the pairs of low- and high-resolution images. The registered image contains padded pixels, while the reference image does not. Thus, to avoid the models from learning misleading information, we used the resulting transformation to reproduce the padding in the reference image.
% all datasets
For all datasets, images were cropped into patches of size $256 \times 256$ pixels in a non-overlapping-fashion.

\subsection{Dataset partitioning}
We here describe (Table \ref{tab:dataset_description}) the number of FOVs, image sizes and dataset partitions used for training, validation and during test time. 
\begin{table}[hbt!]
    \centering
    \resizebox{0.95\columnwidth}{!}{
    \begin{tabular}{c@{\hskip .1in}c@{\hskip .1in}c@{\hskip .1in}c@{\hskip .1in}c@{\hskip .1in}c}
    \toprule
     Dataset &  \# FOVs & Orig. image size (px) & Train & Validation & Test \\
    \midrule
    Microtubules & 104 & $2560 \times 2560$ & 1272 & 89 & 265 \\
    Mitochondria & 345 &  $600 \times 600$ & 2646 & 153 & 306 \\
    Synapses & 24 & $550 \times 550 \times 20$ & 1198 & 56 & 112  \\
    Zebrafish & 20 & $512 \times 512$ & 3600 (72) & 200 (4) & 200 (4)\\
  \bottomrule

    \end{tabular}
    }
    \caption{\textbf{Dataset description and pre-processing}. For each dataset, we report the number of fields of view (FOVs), the image sizes, as well as the number of FOVs for the train, validation and test partitions. For the zebrafish dataset, the same sample is consecutively captured 50 times, exhibiting different noise realizations. Thus, we report in parenthesis the number of different sub-FOVs (after dividing the original into patches) before using all noise realizations.}
    \label{tab:dataset_description}
\end{table}

\subsection{Diversity}
In Table \ref{tab:dataset_diversity} we highlight the dimensions along which the tested datasets vary. 
\begin{table}[hbt!]
    \centering
    \begin{center}
    \resizebox{0.95\columnwidth}{!}{%
  \begin{tabular}{c@{\hskip .1in}c@{\hskip .1in}c@{\hskip .1in}c@{\hskip .1in}c@{\hskip .1in}c} 
    \toprule
     \# & Sample type & Imaging type & Condition & Raw  & Ground Truth\\
    \midrule
    1 & Microtubules & STED & Fixed  & Low-light dose & High-light dose \\
    2 & Mitochondria  &  STED & Living & Low-light dose & High-light dose \\
    3 & Synapses  &  Confocal & Fixed & Confocal & Super-resolution \\
    4 & Zebrafish  & Confocal & Fixed & Single images & Avg. of 50 images \\
  \bottomrule

    \end{tabular}
    }
    \end{center}
    \caption{\textbf{Differences between denoising datasets.} We test the denoising performance using four diverse datasets. These datasets vary along the sample and imaging type, the cell condition (fixed vs. live cells), as well as how the raw and ground truth data were generated.}
    \label{tab:dataset_diversity}
\end{table}

\section{Additional quality control metrics}
\label{SI-additional-metrics}

The mean absolute error (MAE) between the ground truth image $y$ and reconstructed image $\hat{y}$  captures the general offset in pixel values and is calculated as:
\begin{equation}
    MAE(y, \hat{y}) = |y - \hat{y}|.
\end{equation}

The Normalized Root Mean Square Error (NRMSE) compares the pixel values of the reconstruction to the ground truth image. NRMSE normalizes the Root MSE to account for the scale of the data, making it an scalar quantity that is easier to interpret. 

\begin{equation}
    NRMSE(y, \hat{y}) =\frac{\sqrt{MSE(y, \hat{y})}}{\overline{y}}
\end{equation}
Lower NRMSE values indicate a higher correspondence between the ground truth and reconstruction. 

The peak signal-to-noise ratio (PSNR) quantifies the quality of reconstructed images using a logarithmic measure of the peak error (mean squared error, MSE) between $y$ and $\hat{y}$. The PSNR value is expressed in decibels (dB), which logarithmically measures the ratio between the maximum possible pixel value $L$ of the images (here $L=255$) and the MSE:

\begin{equation}
    PSNR(y, \hat{y}) =10 log_{10}\frac{L^2}{MSE}
\end{equation}

Higher PSNR values indicate better image quality, suggesting that the reconstructed image is closer to the original image.

The structural similarity index measure (SSIM) \cite{wang_image_2004} was designed to improve PSNR or MAE by also incorporating differences in luminance $l(y, \hat{y})$, contrast $c(y, \hat{y})$, and structural information $s(y, \hat{y})$. 
The SSIM is defined as: 
\begin{equation}
   SSIM(y, \hat{y})=[l(y, \hat{y})]^\alpha \cdot [c(y, \hat{y})]^\beta \cdot [s(y, \hat{y})] ^\gamma
\end{equation}

where $\alpha$, $\beta$, and $\gamma$ define the relative importance of the three components. Here, we set all to 1 to equally weight each component. The SSIM ranges from 0 (structural dissimilarity) to 1 (perfect structural similarity). The multi-scale SSIM (MS-SSIM) additionally evaluates the structural similarity across various scales to capture both fine details and coarse structures \cite{wang_multiscale_2003}. To this aim, the images are iteratively smoothed using a Gaussian low-pass filter and downsampled by a factor of 2. The SSIM is computed at each scale and the final MS-SSIM score is a weighted product of the SSIM scores of each scale. The weights emphasize different scales based on their importance to human perception. The MS-SSIM ranges from 0 (structural dissimilarity) to 1 (perfect structural similarity). 

The \textit{resolution}, as defined by \cite{descloux_parameter-free_2019}, assesses the resolution of individual images based on decorrelation analysis. The core idea is to examine how the frequency components of the image decorrelate as the distance between them increases, in order to determine the point where significant loss of detail occurs, thereby defining the resolution of the image. High-resolution images have more details and, therefore, higher decorrelation between neighboring pixels. To compute the resolution first standard edge apodization is applied to the image to remove high-frequency artifacts. Then the image is Fourier transformed as $I(\textbf{k})$, where $\textbf{k} = [k_x, k_y]$ represent the coordinates in Fourier space. Additionally, the Fourier transform is normalized as $I_n(\textbf{k}) = \frac{I(\textbf{k})}{|I(\textbf{k})|}$. Next, the cross-correlation between $I(\textbf{k})$ and $I_n(\textbf{k})$ is computed using the Pearson correlation and rescaled to a value between 0 and 1. The calculation is repeated but $I_n(\textbf{k})$ is additionally filtered with a binary circular mask of radius $M(\textbf{k};r)$ with $r \in [0, 1]$. We can then compute the correlation coefficient as:
\begin{equation}
    d(r) = \frac{\iint Re\{I(\textbf{k}) I_n(\textbf{k}) M(\textbf{k}; r)\}dk_x dk_y}{\sqrt{\iint |I(\textbf{k})|^2 dk_x dk_y \iint |I_n(\textbf{k})M(\textbf{k}; r)|^2 dk_x dk_y}}
\end{equation}

For differently high-pass filtered images (from weak to very strong filtering) $d(r)$ is computed and the peak position $r_i$ and amplitude $A_i$ are extracted. The resolution is then defined as the maximum peak across $N_g$ high-pass filters as:

\begin{equation}
R = \frac{2 \times pixel size}{max [r_0, \dots, r_{N_g}]}
\end{equation}
Lower values indicate a better resolution, as more fine-grained features are visible.

As the resolution is measured on each image individually, we propose a method for denoising tasks that computes the performance, respectively to the high-resolution data. Specifically, we compute: 
\begin{equation}
    \bar{R} = \frac{R_{\hat{y}}}{R_y}
\end{equation}
where $R_y$ and $R_{\hat{y}}$ refer to the resolution of the high-intensity image $y$ and predicted image $\hat{y}$, respectively. Values close to 1 indicate similar resolution between the high-intensity image and the prediction, i.e. $R_y \approx R_{\hat{y}}$. Values above (resp. below) 1 indicate that the prediction exhibits worse (resp. better) resolution than the ground-truth high-intensity image.

The learned perceptual image patch similarity (LPIPS) \cite{zhang_unreasonable_2018} assesses the perceptual similarity between images. In contrast to PSNR and SSIM, LPIPS compares feature representations extracted from a pre-trained deep neural network (here AlexNet) to assess perceptual similarity, which often aligns more closely with human visual perception. The LPIPS value ranges from 0 (high perceptual similarity) to 1 (low perceptual similarity). 

\section{Benchmark models}
Here, we describe the specific setup and training conditions for each benchmark model.
\begin{itemize}
    \item Noise2Void \cite{krull_noise2void_2019} - We use the TensorFlow implementation from the authors. Epochs: 100, batch size: 32, initial learning rate: $2\mathrm{e-}4$. All other parameters use the default. We use the best-trained state identified by default by Noise2Void.
    \item pix2pix \cite{sun_pix2pix_2022} - We use the implementation from ZeroCostDL4Mic \cite{von_chamier_democratising_2021}. Epochs: 5, batch size: 1, initial learning rate: $2\mathrm{e-}4$.
    \item UNet-RCAN \cite{ebrahimi_deep_2023} - Default settings. Max epochs: 200, initial learning rate: $1\mathrm{e-}4$, batch size: 1. We use the best-trained state identified by UNet-RCAN.
    \item CARE \cite{weigert_content-aware_2018} - We use the implementation from ZeroCostDL4Mic. Epochs: 1000, batch size: 8, initial learning rate: $4\mathrm{e-}4$. We used the best-trained state identified by default by CARE.
\end{itemize}

\section{Versions}
To compute the mean absolute error (MSE) and Pearson correlation, we use NumPy v1.24.4 \cite{harris_array_2020}. The peak signal-to-noise ratio (PSNR) is computed using the scikit-image library v0.19.3 \cite{walt_scikit-image_2014}. The multi-scale structural similarity index measure (MS-SSIM) and learned perceptual image patch similarity (LPIPS) are computed using Torchmetrics v1.3.1. The resolution is computed using the plugin ImageDecorrelationAnalysis \cite{descloux_parameter-free_2019} for ImageJ \cite{schindelin_fiji_2012}.

\section{Averaging across many reconstructions}
\label{SI-averaging}
To improve the performance of the DDPM and remove any noise that was not removed by the denoising process, we employ an averaging strategy. Specifically, we generate several images using the same conditioning input but different inference runs. We consistently observe an increase in performance across several metrics when averaging, except for LPIPS (see Fig. \ref{SI-effect-avg-all-metrics}), and in some cases resolution (see DDPM vs. DDPM-avg for microtubule and synapse in Fig. \ref{tab: resolution}). This might be explained by the smoothing effect of averaging which removes fine-grained structures. Note that this fine-grained structure is not always desirable to keep in the image and might also indicate noise. Moreover, we observe the performance saturating with approximately 10 averaged samples.

\begin{figure}[hbt!]
  \centering
  \includegraphics[width=0.258\textwidth]{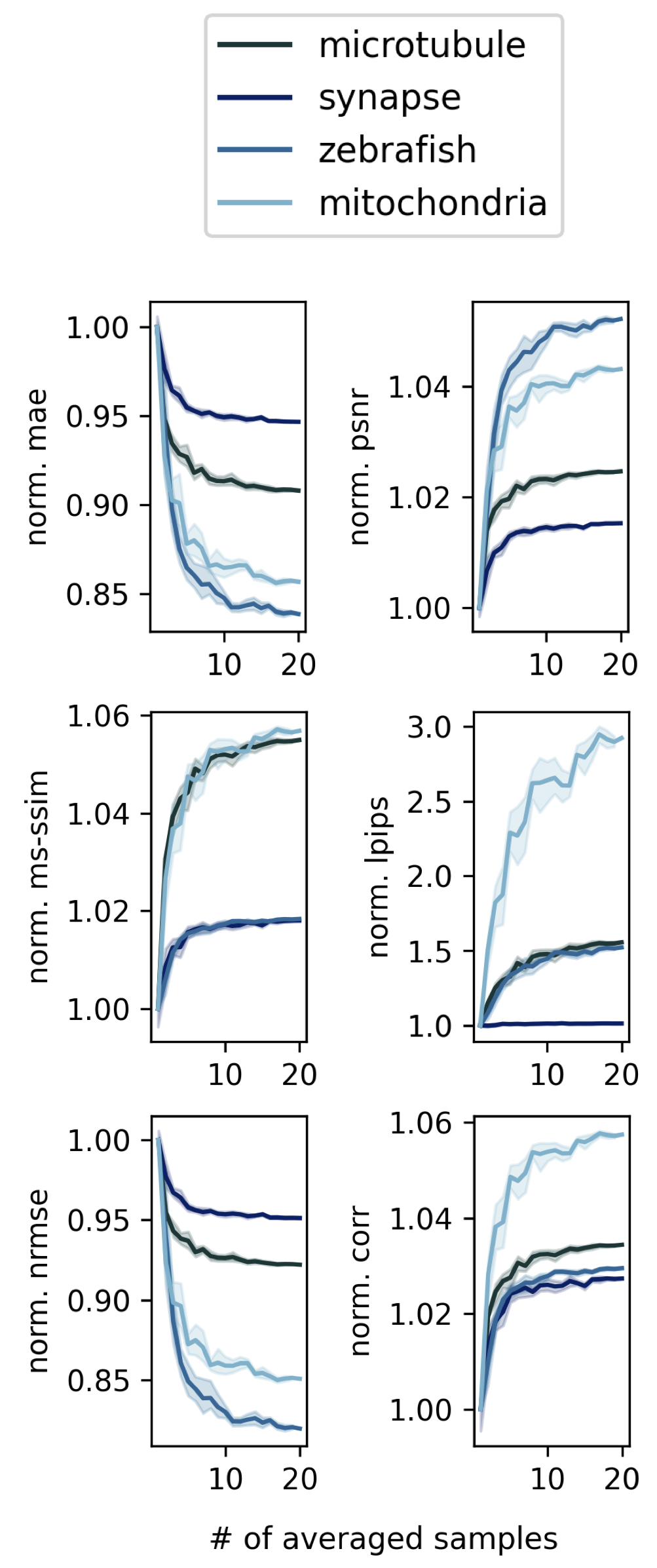}
  \caption{\textbf{Averaging across samples improves the performance for most metrics for the DDPM}. We repeatedly predict a denoised image using the same low-intensity conditioning input but different initial noise. We compute the mean image across different numbers of reconstructions. Average performance is shown in bold, and the translucent ban indicates the standard deviation.} % std across different subsets of runs averaged
  \label{SI-effect-avg-all-metrics}
\end{figure}
\newpage

\subsection{Uncertainty maps}

We benefit from the above-described repeated sampling strategy to enhance the interpretability of the model. In particular, repeated sampling is valuable as it captures the variability of the model, thus reflecting its uncertainty in restoring certain areas of the image. After the model performs inference multiple times with the same conditioning input but different inference runs, we approximate the uncertainty based on the pixel-wise standard deviation (Eq. \ref{eq: std-uncertainty}), and on the pixel-wise entropy (Eq. \ref{eq: entropy-uncertainty}) across the different model outputs. In principle, it is also possible to compute uncertainty in a more abstract-fashion using the latent representations of the predicted image, i.e. in the $\mathcal{H}$-space of diffusion models, which we leave for future work.

Uncertainty maps provide us with a tool to verify that the model has learnt to restore regions in the image acc. For instance, one would expect complex and inherently ambiguous areas such as edges, to be predicted with a high uncertainty, otherwise suggesting over-fitting. Likewise, simple and smooth regions are expected to be predicted with low uncertainty, otherwise a sign of potential under-fitting. Additionally, if one were to collect additional data to refine the model, uncertainty maps can pinpoint the sub-structures that the current model struggles with, thus enabling a more informed data collection.

Additional to elucidating potential areas of improvement in the model, uncertainty maps can also be useful during the post-processing of the data, by informing about regions that could require further visual inspection or manual processing.

Given $N$ repeated predictions $\{\hat{y}^1, ..., \hat{y}^N\}$ from the same noisy image, we compute the standard deviation-based uncertainty map as:
\begin{equation}
\label{eq: std-uncertainty}
    \sqrt{\frac{\sum_{i=1}^{N}\left({\hat{y}^i - \bar{\hat{y}}}\right)^2}{255^2N}},
\end{equation}

where N = 15 is the number of times we repeat the sampling, $\bar{\hat{y}}$ is the average of the multiple predicted samples, and $255^2$ is a normalization factor to constrain values between 0 and 1. We illustrate several examples in Fig. \ref{fig: SI-uncertainy-maps}.

Whereas the entropy-based uncertainty map is $S = (s_{jk})_{1 \leq j \leq 256, 1 \leq k \leq 256}$: 
% write formula for entropy-based uncertainty?
\begin{equation}
\label{eq: entropy-uncertainty}
    s_{jk} = - \sum_{m=1}^{M} p_m \log p_m,
\end{equation}

where $M$ is the number of unique pixel values at location $(j, k)$ among the single image predictions, and $p_m$ is the probability of the $m$-th unique pixel value at location $(j, k)$.

\newpage
\begin{figure}[hbt!]
  \centering
  \includegraphics[width=0.5\textwidth]{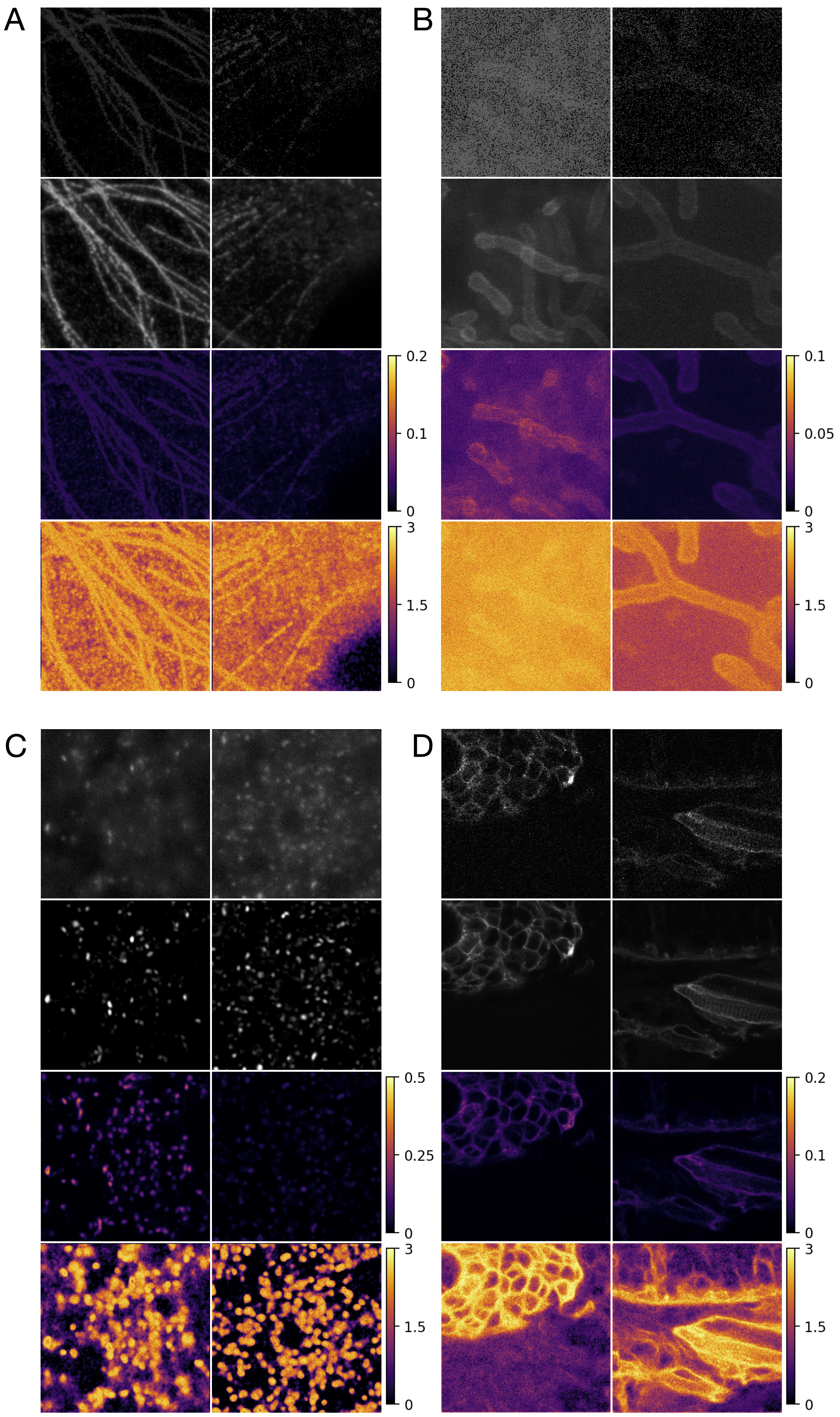}
  \caption{\textbf{Uncertainty maps based on repeated sampling strategy with DDPM.} For each dataset (\textbf{A}: microtubules, \textbf{B}: mitochondria, \textbf{C}: synapse, \textbf{D}: zebrafish), we show a subfigure with two low- (first row) and high- (second row) resolution images, and the resulting uncertainty maps, based on pixel-wise standard deviation (second row) or on entropy (third row). Note that for better visibility, the standard deviation-based uncertainty range is different for every dataset. Likewise, the pixel range was adjusted for the noisy images of mitochondria and microtubules.} 
  \label{fig: SI-uncertainy-maps}
\end{figure}
% alternative title: repeated sampling on DDPMs reveals regions of uncertainty in predictions

When computing uncertainty as the pixel-wise standard deviation, we find that many high uncertainty regions correspond to the brighter areas of the low-resolution images. This might be due to small variations in intensity being amplified when calculating their difference. Another factor that could explain the higher uncertainty in bright regions is the complex structure underlying these areas, making their reconstruction more challenging for the model. Additionally, the model could be over-relying on these bright features to reconstruct the multiple samples, which would indicate a bias in how the model handles intensity features. Moreover, the model shows the highest uncertainty for the synapse dataset (see Fig. \ref{fig: SI-uncertainy-maps}C), whereas the mitochondria dataset has the lowest uncertainty values (see Fig. \ref{fig: SI-uncertainy-maps}B. In particular, for mitochondria, the model is most uncertain in predicting the membrane, an area which is inherently ambiguous in the noisy data (see Fig. \ref{fig: SI-uncertainy-maps}B).

In contrast, uncertainty regions for the entropy-based formulation go beyond bright areas, and also include very noisy background regions. Combined with the previous observations, this can be interpreted as the predicted pixel intensities being uniformly distributed in a narrow range of values, which is a positive feature given the absence of complex structures on those regions, and namely the case for the background in the microtubules and the mitochondria datasets (see Fig. \ref{fig: SI-uncertainy-maps}A, B). Furthermore, on the zebrafish images, we observe high uncertainty also in regions with visibly fine-grained details in the high-resolution image, that are ambiguous in the low-resolution image due to overlaid noise (see Fig. \ref{fig: SI-uncertainy-maps}D). Thus, the model has not learnt to restore such small structures from noisy images.

In both uncertainty formulations, smooth regions in the noisy images are characterized by high-confidence values in the uncertainty maps, which reflects the model's ability to reliably predict non-complex regions.

% could also add the results for entropy-based uncertainty (useful for multiple plausible outputs for the same conditioning input)

\clearpage

\section{Results on additional metrics}
\label{sec: SI-results-additional-metrics}
Additionally to the results reported in the main text, we include additional metrics here (Fig. \ref{fig:SI-supp-metrics}). Specifically, we report the performance of all models on the  NRMSE and Pearson correlation for the internal (Table \ref{tab:supp-metrics-internal}) and external (Table \ref{tab:supp-metrics-external}) datasets. 

\begin{figure}[hbt!]
  \centering
  \includegraphics[width=0.5\textwidth]{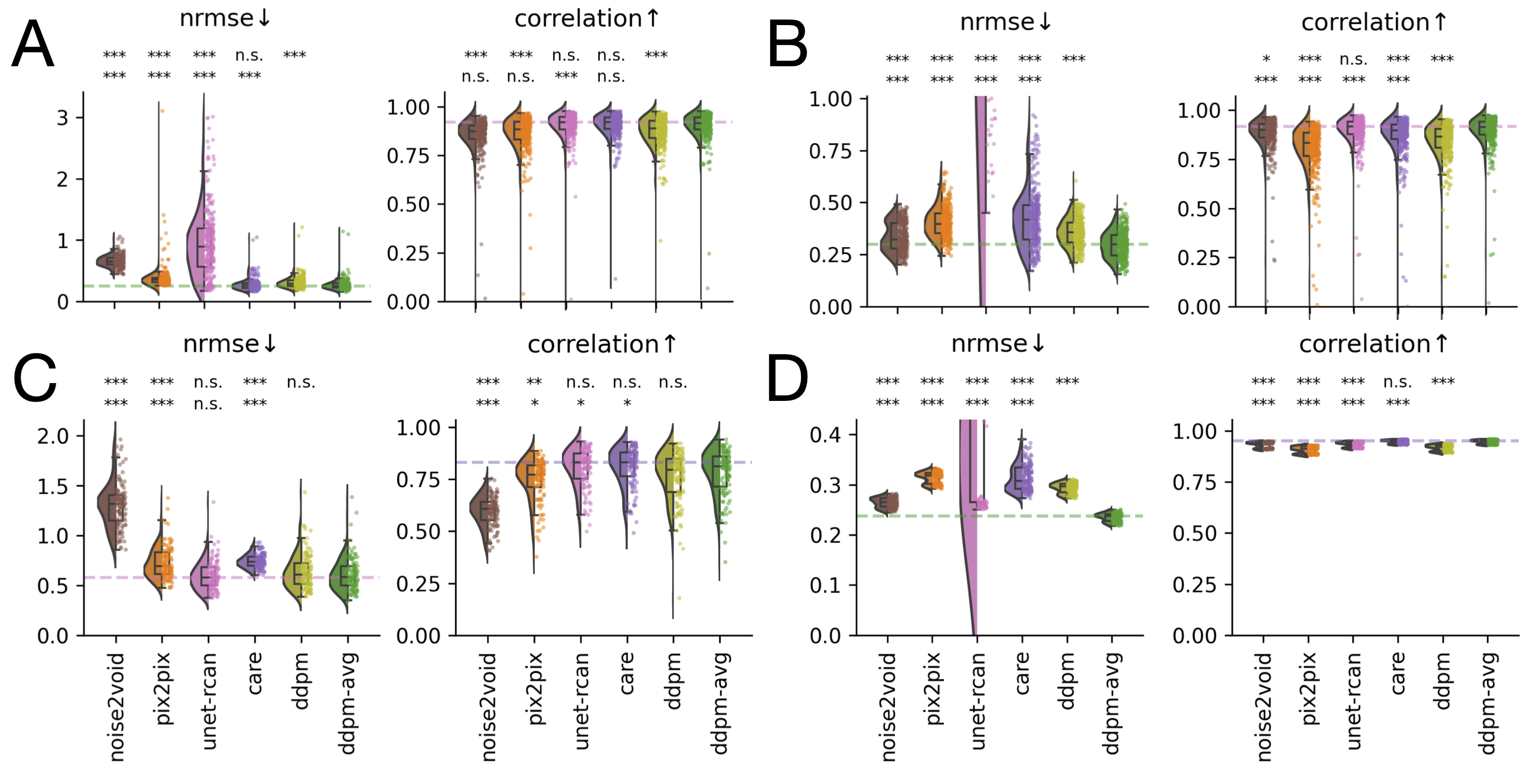}
  \caption{\textbf{Conditioned DDPMs outperform several previous methods in denoising STED and confocal images.} Performance comparison on additional metrics between our method and several previously proposed benchmark models for the microtubule (\textbf{A}), mitochondria (\textbf{B}), synapse (\textbf{C}), and zebrafish (\textbf{D}) datasets. We indicate the median of the best-performing model for each metric as a dashed line in the respective color. Mood's median test was used to compute statistical significance, ***: $p < .001$, **: $p < .01$, *: $p < .05$, otherwise not significant. In the upper (resp. lower) row, significance is indicated for the DDPM-avg (resp. DDPM).} % show in the supp mat how other metrics evolve with averaging?
  \label{fig:SI-supp-metrics}
\end{figure}

\begin{table}[hbt!]
  \centering
  \begin{center}
  \resizebox{0.9\columnwidth}{!}{%
  \begin{tabular}{c@{\hskip .1in}c@{\hskip .1in}c@{\hskip .1in}c@{\hskip .1in}c@{\hskip .1in}c} 
    \toprule
           & Microtubule & & Mitochondria & \\ 
     Model & NRMSE & Corr. & NRMSE  & Corr. \\
    \midrule
    Raw & $0.99$ & $0.46$  & $0.97$ & $0.40$ \\
    Noise2Void  &  $0.65$ & $0.87$ & $0.32$ & $0.90$ \\
    Pix2pix  &  $0.35$ & $0.88$ & $0.40$ & $0.83$ \\
    UNet-RCAN  & $0.90$ & $\textbf{0.92}$ & $3.76$ & $\textbf{0.92}$ \\
    CARE  & $0.26$ & $\textbf{0.92}$ & $0.42$ & $0.90$ \\
    \hline
    \textbf{DDPM}  &
    $0.29$ &
    $0.89$ &
    $0.36$ &
    $0.87$ \\
    \textbf{DDPM-avg} & $\textbf{0.25}$ & $\textbf{0.92}$ & $\textbf{0.30}$ & $\textbf{0.92}$ \\
  \bottomrule
  \end{tabular}%
  }
\end{center}
  \caption{\textbf{Benchmarking the conditioned DDPM with additional metrics.} We report the median value of additional performance metrics, NRMSE (the lower the better) and Pearson correlation (the higher the better), across our two novel datasets.}
  \label{tab:supp-metrics-internal}
\end{table}

\begin{table}[hbt!]
  \centering
  \begin{center}
  \resizebox{0.8\columnwidth}{!}{%
  \begin{tabular}{c@{\hskip .1in}c@{\hskip .1in}c@{\hskip .1in}c@{\hskip .1in}c@{\hskip .1in}c} 
    \toprule
           & Synapse & & Zebrafish &\\ 
     Model & NRMSE & Corr. & NRMSE  & Corr.  \\
    \midrule
    Raw & $1.33$ & $0.60$ & $0.70$ & $0.74$ \\
    Noise2Void  &  $1.32$ & $0.61$ & $0.27$ & $0.94$ \\
    Pix2pix  & $0.69$ & $0.77$ & $0.32$ & $0.91$ \\
    UNet-RCAN  & $\textbf{0.58}$ & $\textbf{0.83}$ & $0.55$ & $0.94$ \\
    CARE  & $0.74$ & $\textbf{0.83}$ & $0.31$ & $\textbf{0.95}$\\
    \hline
    \textbf{DDPM}  & $0.61$ & $0.80$ & $0.30$ & $0.92$ \\
    \textbf{DDPM-avg} & $\textbf{0.58}$ & $0.81$ & $\textbf{0.24}$ & $\textbf{0.95}$ \\
  \bottomrule
  \end{tabular}%
  }
\end{center}
  \caption{\textbf{Benchmarking the conditioned DDPM with additional metrics.} Perfomance evaluation with NRMSE (the lower the better) and Pearson correlation (the higher the better) across the two external datasets.}
  \label{tab:supp-metrics-external}
\end{table}
\newpage
\subsection{Reconstruction resolution}
Additionally to the above-reported performance metrics, we also compute the resolution as proposed by Descloux et al. \cite{descloux_parameter-free_2019}, as well as the resolution of the reconstruction scaled by that of the ground truth (resolution ratio; see Table \ref{tab: resolution}). The resolution indicates the scale of the smallest fine-grained structure visible in the image. We observe that pix2pix performs best for the fixed-cell microtubules and zebrafish datasets, Noise2Void on the synapse dataset, and UNet-RCAN on the live-cell mitochondria dataset. In particular, the resolution for the low-resolution images (raw) is lower than the high-resolution images (GT), suggesting the presence of artifacts, which is misleading for the evaluation of this metric for the synapse dataset. Note that all other evaluation metrics rate these methods poorly on the respective datasets. However, these metrics mostly rely on some form of pixel-wise error, whereas the resolution is based on cross-correlations within the image in the frequency domain. However, we observe that the resolution often picks up high-frequency noise in the data which wrongly improves the results.

\begin{table}[hbt!]
  \resizebox{\columnwidth}{!}{%
\begin{tabular}{c@{\hskip .1in}c@{\hskip .1in}c@{\hskip .1in}c@{\hskip .1in}c} \\ \hline
\toprule
    & Microtubule & Mitochondria & Synapse & Zebrafish \\
     Model & $r$ / $r$ ratio & $r$ / $r$ ratio & $r$ / $r$ ratio & $r$ / $r$ ratio \\
    \midrule
Raw     & $128.60$ / $1.3$ & $3563.64$ / $11.91$ & $143.14$ / $0.49$ & $5297.4$ / $6.82$ \\
GT         &  $98.80$ /  $1.00$  & $299.24$ / $1.00$ &  $293.33$  /  $1.00$  & $776.70$  /  $1.00$  \\
\midrule
Noise2Void &  $107.85$  /  $1.09$  & $111.36$ / $0.37$ & $\textbf{147.04}$ / $0.50$ & $1141.8$ / $1.47$ \\
Pix2pix    & $\textbf{88.45}$  / $0.90$ & $149.62$ / $0.50$ & $230.58$ / $0.79$ &  $\textbf{730.05}$ /  $0.94$\\
UNet-RCAN  & $118.35$ /  $1.20$ & $\textbf{76.72}$ / $0.27$ & $385.88$ / $1.32$  &  $1031.70$  / $1.33$  \\
CARE       & $119.75$ / $1.21$ & $137.54$ / $0.46$ & $363.20$ / $1.24$ &  $772.35$ /  $0.99$  \\
\midrule
DDPM       &  $97.6$ / $0.99$ & $177.38$ / $0.59$ & $330.18$ / $1.13$  &  $831.60$  /  $1.07$ \\
DDPM-avg & $115.28$ / $1.17$ & $110.74$ / $0.37$ & $363.20$ / $1.24$ & $777.75$ / $1.00$ \\
 \bottomrule
\end{tabular}%
}
\caption{\textbf{Resolution across models and datasets.} We report the median of image resolution in nm, and the resolution ratio with respect to ground-truth (GT) resolution.}
\label{tab: resolution}
\end{table}
\clearpage

\section{Model architecture}
\label{SI-architecture}

\begin{figure}[hbt!]
  \centering
  \includegraphics[width=0.5\textwidth]{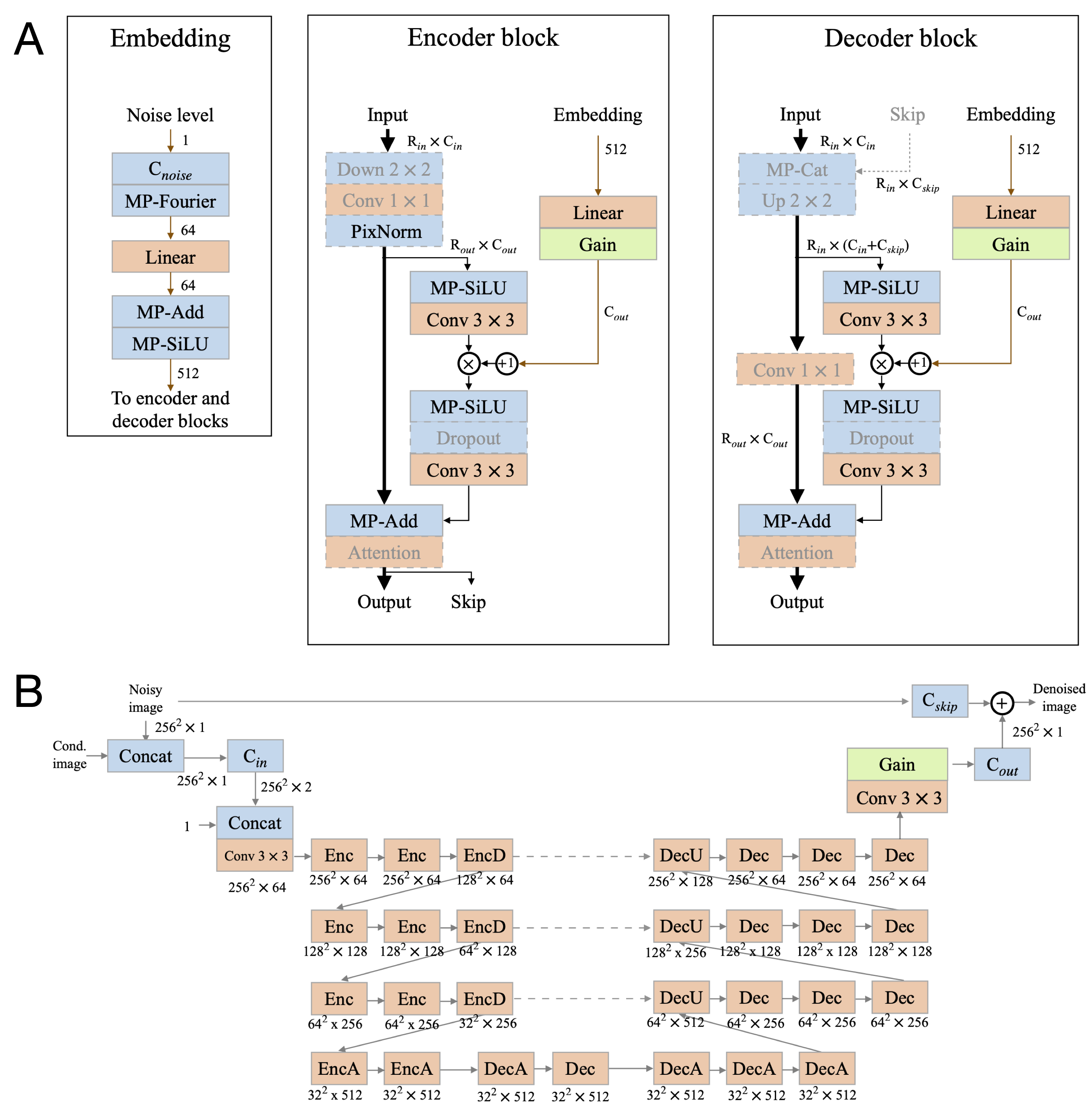}
  \caption{\textbf{U-Net architecture.} Adapted from Karras et al. \cite{karras_analyzing_2024}. \textbf{A}) We depict the three main parts of the U-Net model: an auxiliary embedding network that conditions the U-Net according to the noise level, encoder blocks that gradually decrease the resolution of the image, and decoding blocks that gradually increase it. \textbf{B}) The network receives as input the noisy image concatenated to the conditioning image (low-resolution image in our case). This is then processed by the encoder and decoder blocks following the main path (solid arrows), that additionaly communicate between them via skip connections (dashed arrows). \textit{EncD} and \textit{EncA} are encoder blocks that include downsampling and self-attention, respectively. This is analogous to decoder blocks \textit{DecD} and \textit{DecA}. $c_{in}$, $c_{out}$, $c_{skip}$ are constants that depend on the noise level. MP stands for Magnitude-Preserving. Layers are color-coded as follows: green - parameters are learned, clay - parameters are learned with \textit{forced weight normalization}, blue - function is fixed, dashed contour - not always present.}
  \label{fig: SI-supp-architecture}
\end{figure}

  \subsection{Timestep embedding}
  
  As in \cite{karras_analyzing_2024}, we replace ADM’s original timestep embedding layer, and instead embed the noise level information as Fourier features:
  \begin{equation}
  \label{eq: SI-Fourier_emb}
  MPFourier(a) =
    \left[ 
      \begin{matrix}\sqrt{2} cos(2\pi(f_1 a + \varphi_1))\\
      \sqrt{2} cos(2\pi(f_2 a + \varphi_2))\\
      \vdots \\ 
      \sqrt{2} cos(2\pi(f_N a + \varphi_N))\\ \end{matrix} \right],
  \end{equation}
  
  where $f_i \sim \mathcal{N}(0,1)$, $\varphi \sim \mathcal{U}(0,1)$, and $a = \bar{\alpha}_t$ is a scalar defined as a function of the noise level $t$ and the variance schedule. In the feature vector, $\sqrt{2}$ is the scaling factor that enables magnitude preservation, followed by a linear transformation (as shown in Fig. \ref{fig: SI-supp-architecture}A) with learnable parameters, a magnitude-preserving sum operator, and a magnitude-preserving SiLU non-linearlity.

\end{document}